\documentclass[11pt,a4paper]{article}%

\usepackage{jheppub}
\usepackage{epsf}

\def\be{\begin{equation}}
\def\eqn#1{\be\label{#1}}
\def\ee{\end{equation}}

\def\bb {\begin {eqnarray}}
\def\eqnn#1{\bb\label{#1}}

\def\eea{\end{eqnarray}}

\newcommand{\eqna}[1]{\begin{subequations} \label{#1}
\begin{eqnarray}}
\def\eena{\end{eqnarray}
\end{subequations}}

\def\md{\medskip}
\def\nd{\end{document}}

\def\veps{\varepsilon}

\def\tablerule{\noalign{\hrule}}
\def\has{{\hat\alpha}}
\def\ta{{\tilde\alpha}}
\def\tchi{{\tilde\chi}}
\def\hchi{{\hat\chi}}

\def\hh{h}
\def\tk{{\tilde k}}

\def\nn{\nonumber}
\def\nt{\noindent}
\def\nl{\hfill\break}

\def\np{\vfill\eject}

\def\han{{\textstyle\frac{p+q-2}{2}}}
\def\hal{{\textstyle\frac{p+q}{2}}}
\def\hann{{\textstyle{n\over2}}}
\def\haq{{\textstyle{q\over2}}}
\def\ham{{\textstyle{m\over2}}}
\def\hap{{\textstyle{p\over2}}}
\def\hapq{{\textstyle{p+q\over4}}}

\def\hel{{\textstyle{11\over2}}}

\input epsf.tex
\newcount\figno
\def\fig#1#2#3{
\par\begingroup\parindent=0pt\leftskip=1cm\rightskip=1cm\parindent=0pt
\baselineskip=11pt \global\advance\figno by 1 
\epsfxsize=#3 \centerline{\epsfbox{#2}} \vskip 12pt
#1\par
\endgroup\par}
\def\figlabel#1{\xdef#1{\the\figno}}
\def\encadremath#1{\vbox{\hrule\hbox{\vrule\kern8pt\vbox{\kern8pt
\hbox{$\displaystyle #1$}\kern8pt} \kern8pt\vrule}\hrule}}

  \def\tV{{\tilde V}}
\def\tcn{{\tilde{\cal N}}}

\def\bu{\noindent $\bullet~$}
\def\rank{{\rm rank}\,}
\def\deg{{\rm deg}\,}
\def\downcirc#1{\mathop{\circ}\limits_{#1}}
\def\riga{-\kern-4pt - \kern-4pt -}
\font\fat=cmsy10 scaled\magstep5

\def\Bbullet{\raise-3pt\hbox{\fat\char"0F}}

\def\Box{{\square}}

\def\down{\raise1.5pt\hbox{$\phantom{a}_2$}\downarrow}

\def\downa{\raise1.5pt\hbox{$\phantom{a}_{2\atop m_2}$}\downarrow}

\def\llr{\longrightarrow}

\def\({\left(}
\def\){\right)}
\def\eps{\epsilon}
 
\def\lra{\longrightarrow}
\def\llra{\longleftrightarrow}

\def\dia{{$\diamondsuit$}}

\def\ha{{\textstyle{\frac{1}{2}}}}

\def\bbz{\mathbb{Z}}
\def\bbc{\mathbb{C}}
\def\bac{\bbc} 
\def\bbr{\mathbb{R}}
\def\bbq{\mathbb{Q}}
\def\bbo{\mathbb{O}}
\def\bbn{\mathbb{N}}
\def\a{\alpha}
\def\b{\beta}
\def\d{\delta}

\def\vr{\vert}

\def\D{{\Delta}}

\def\ed{\end{document}}

\def\ca{{\cal A}}  \def\cc{{\cal C}}
\def\cd{{\cal D}} \def\ce{{\cal E}} \def\cf{{\cal F}}
\def\cg{{\cal G}} \def\ch{{\cal H}} 
 \def\ck{{\cal K}} 
\def\cm{{\cal M}} \def\cn{{\cal N}} 
\def\cp{{\cal P}}  
 \def\ct{{\cal T}}

\def\ido{intertwining differential operator}
\def\idos{intertwining differential operators}

\def\L{\Lambda}
\def\r{\rho}

\title{Invariant Differential Operators for Non-Compact Lie Algebras
Parabolically Related to Conformal Lie Algebras}

\author[]{V.K. Dobrev}

\affiliation[]{Theory Division, Department of Physics, CERN,\\ CH-1211 Geneva 23, Switzerland,  \\
Vladimir.Dobrev@cern.ch\\
permanent address: \\
Institute for Nuclear Research and Nuclear Energy,\\
Bulgarian Academy of Sciences,
\\ Tsarigradsko Chaussee 72, BG-1784 Sofia, Bulgaria}

\abstract{In the present paper we continue the project of systematic
construction of invariant differential operators for non-compact
semisimple Lie groups. Our starting points is the class of algebras,
which we call 'conformal Lie algebras' (CLA), which have very
similar properties to the conformal algebras of  Minkowski
space-time, though our aim is to go beyond this class in a natural
way.  For this we introduce the new notion of ~{\it parabolic
relation}~ between two non-compact semisimple Lie algebras $\cg$ and
$\cg'$ that have the same complexification and possess maximal
parabolic subalgebras with the same complexification. Thus, we
consider
 the exceptional algebra   ~$E_{7(7)}$ which is parabolically related to the CLA ~$E_{7(-25)}\,$,
the parabolic subalgebras including ~$E_{6(6)}$ and ~$E_{6(-26)}$.
Other interesting examples are the orthogonal algebras ~$so(p,q)$~ all of which are parabolically
related to the conformal  algebra ~$so(n,2)$~ with ~$p+q=n+2$, the parabolic subalgebras including
the Lorentz subalgebra ~$so(n-1,1)$~ and its analogs ~$so(p-1,q-1)$.
We consider also ~$E_{6(6)}$~ and ~$E_{6(2)}$~
which are parabolically related to the hermitian symmetric case ~$E_{6(-14)}\,$,
the parabolic subalgebras including real forms of $sl(6)$.\\
We also give a formula for the number of representations in the main
multiplets valid for CLAs and all algebras that are parabolically
related to them. In all considered cases we give the main multiplets
of indecomposable elementary representations including the necessary
data for all relevant invariant differential operators. In the case
of $so(p,q)$ we give   also the  reduced multiplets. We should
stress that the multiplets are given in the most economic way in
pairs of ~{\it shadow fields}.  Furthermore we should stress that
the classification of all invariant differential operators includes
as special cases all possible ~{\it conservation laws}~ and ~{\it
conserved currents}, unitary or not.}


\begin{document}
\maketitle

\np

\section{Introduction}

 {\it Invariant differential operators} play very important role in
the description of physical symmetries - starting from the early
occurrences in the Maxwell, d'Allembert, Dirac, equations, (for
more examples cf., e.g., \cite{Corn}),  to the
latest applications of (super-)differential operators in conformal
field theory, supergravity and string theory (for    reviews, cf. e.g.,
\cite{Mald},\cite{Ter}.

For example,  applications of invariant differential operators in
supersymmetry  involved the study of multiplets, superfields and
supercurrents \cite{FerrWessZumFay,OgSok}, of harmonic superspaces
\cite{Sokat,GIKOS}, of auxiliary fields of supergravity
\cite{FrvNFerr}, on the coupling of supersymmetric Yang-Mills
theories to supergravity \cite{CrSchFerr}, twistor formulation of
superstrings \cite{Delduc}, Landau-Ginzburg description of $N=2$
minimal models \cite{Witten}, in various other applications to
superstrings and supergravity \cite{CerDAFerr,AFT,AGMOO}.

Invariant differential operators played important role in the
group-theoretical approach to  conformal  field theory
\cite{DPPT,DMPPT,TMP}, e.g.,  in the derivation of operator product
expansion of two scalar fields.

Invariant super-differential operators were crucial in the
derivation of the classification of positive energy unitary
irreducible representations of extended conformal supersymmetry in
4D \cite{DobPet}, later in 3D \& 5D \cite{Min}, in 6D
\cite{Min,Dob6D}, (see also \cite{Vara}), then for the derivation of
the character formulae in 2D \cite{DobN2char}. Later applications
include
\cite{APSW,ABCDFFM,FerrMal,FerrFr,EdSok,DNW,FerrSok,FaKoRi,KiMaMi,GuLuMi,HofMal,Mizo}.

Special mentioning requires the applications of exceptional groups,
cf.
\cite{Biedenharn,Gursey,Ferrara,Nicolai,FerrKaMa,DufFer,BCCS,Kallosh,BiFerr,CedPal,Brink,GunPav,CCM,Marrani,GSS},
since they play important role in the present paper. Exceptional
groups recently appeared also as symmetries of "Freudenthal dual"
Lagrangians, as investigated, e.g., in \cite{BDFM}.

Finally, among our motivations are the mathematical developments -
see the relevant mathematical references:
\cite{Har,BGG,War,Lan,Zhea,Kosa,Wolfa,Wolfb,KnZu,SpVo,Vog,EHW,BOO,TruVar,KacWak,Kob,Eas,Knab,KacRoWa,Kosb,BaWa},
and others throughout the text.

\md

Thus, it is important for the applications in physics to study
systematically such operators.  In a recent paper \cite{Dobinv} we
started the systematic explicit construction of invariant
differential operators. We gave an explicit description of the
building blocks, namely, the {\it parabolic subgroups and
subalgebras} from which the necessary representations are induced.
Thus we have set the stage for study of different non-compact
groups.

Since the study and description of detailed classification should be
done group by group we had to decide which groups to study first. A natural
choice would be non-compact groups that have ~{\it discrete series}~
of representations. By the Harish-Chandra criterion \cite{HC} these are groups
where holds:
$$ \rank G = \rank K, $$
where $K$ is the {\it maximal compact subgroup} of the non-compact group
$G$. Another formulation is to say that the Lie algebra $\cg$ of $G$
has a compact Cartan subalgebra.

\noindent {\it Example:}  the groups {\it $SO(p,q)$} have discrete series,
{\it except} when both $p,q$ are {\it odd} numbers.

This class is still rather big, thus, we decided to consider a subclass,
namely, the class of {\it Hermitian symmetric spaces}. The practical criterion is that in
these cases, the {\it maximal compact subalgebra} $\ck$ is of the form:
\eqn{hermss} \ck ~=~ so(2) \oplus \ck'    \ .\ee
The Lie algebras from this class are:
\eqn{herm} so(n,2), ~~sp(n,R), ~~su(m,n),  ~~so^*(2n), ~~E_{6(-14)}\,,
~~E_{7(-25)}  \ee These groups/algebras have {\it highest/lowest weight
representations}, and relatedly {{\it {\it holomorphic} discrete series
representations}.

The most widely used of these algebras are the {\it conformal  algebras}
~$so(n,2)$~ in $n$-dimensional Minkowski space-time. In that case,
there is a maximal {{\it Bruhat decomposition} \cite{Bru}:
\eqnn{bruc} &&so(n,2) ~=~ \cp\,\oplus\,\tcn ~=~ \cm\,\oplus\,\ca\,\oplus\,\cn\,
\oplus\,\tcn \ ,\\ && \cm ~=~ so(n-1,1) \ , ~~\dim\ca=1, ~~\dim \cn =
\dim\tcn = n \nn\eea
that has direct physical meaning, namely,
 ~$so(n-1,1)$~ is the {\it Lorentz algebra} of
$n$-dimensional Minkowski space-time, the subalgebra ~$\ca ~=~
so(1,1)$~ represents the {\it dilatations}, the conjugated subalgebras
~$\cn\,$, $\tcn\,$~ are the algebras of {\it translations}, and {\it special
conformal transformations}, both being isomorphic to $n$-dimensional
Minkowski space-time. The subalgebra ~$\cp = \cm\,\oplus\,\ca\,\oplus\,\cn\,$
 ($\cong \cm\,\oplus\,\ca\,\oplus\,\tcn\,$) is  a maximal parabolic
 subalgebra.\footnote{The precise general definition is given in Section 2.}

There are other special  features which are important. In
particular, the complexification of the maximal compact subgroup
is isomorphic to  the complexification of the first two factors of the
Bruhat decomposition: \eqn{relkm}  \ck^\bac ~=~ so(n,\bbc)  \oplus so(2,\bbc)
~\cong~ so(n-1,1)^\bac  \oplus so(1,1)^\bac = \cm^\bac \oplus\ca^\bac
\ .\ee
 In particular, the coincidence of the complexification of the
semi-simple subalgebras:
\eqn{relkmc}
\ck'^\bac ~=~ \cm^\bac \ee   means
that the sets of finite-dimensional (nonunitary) representations of
~$\cm$~ are in 1-to-1 correspondence with the finite-dimensional
(unitary) representations of ~$so(n)$. The latter leads to the fact
that the corresponding induced representations
 are representations of finite $\ck$-type \cite{HC}.

It turns out  that some of the hermitian-symmetric algebras  share
the above-mentioned special properties of ~$so(n,2)$. That is why,
in view of applications to physics, these algebras   should be called '{\it conformal
Lie algebras}' (CLA), (or groups).

  This subclass  consists of:
\eqn{confc} so(n,2), ~~sp(n,\bbr), ~~su(n,n),  ~~so^*(4n),
~~E_{7(-25)}  \ee   the corresponding analogs of Minkowski
space-time $V$ being:
\eqn{nherm} \bbr^{n-1,1}, ~~{\rm Sym}(n,\bbr), ~~{\rm Herm}(n,\bbc),
~~{\rm Herm}(n,\bbq), ~~ {\rm Herm}(3,\bbo) \ .\ee

The corresponding groups are also called '{\it Hermitian symmetric spaces
of tube type}' \cite{FaKo}.
  The same class was identified from different considerations in \cite{Guna}
    called there '{\it conformal groups  of simple Jordan
    algebras}'. In fact, the relation between Jordan algebras
and division algebras was known long time ago.   Our class  was
identified from still different considerations also in
\cite{Mackder}    where they    were called '{\it simple space-time
symmetries generalizing conformal symmetry}'.   For more references
on Jordan algebras relevant in our approach cf., e.g.,
\cite{Okubo,GunSac,TruBied,GuSiTo,Sierra,CecFerrGir,Cederwall,Hull,Tru,Ramond,Catto,FerrMa,ACCF}.

\md

We have started the study of the above class in the framework of the
present approach in the cases:
~$so(n,2), ~~su(n,n), ~~sp(n,\bbr),  ~~E_{7(-25)}$, in
 \cite{Dobpeds}, \cite{Dobsunn}, \cite{Dobspn},  \cite{Dobeseven}, resp.,
and we have considered also the algebra $E_{6(-14)}$, \cite{Dobesix}.

In the present paper we are mainly interested in non-compact Lie algebras (and groups)
that are  'parabolically' related to the conformally Lie algebras.

\bu {\it Definition:}  ~~~Let ~$\cg,\cg'$~ be two non-compact semisimple Lie algebras
with the same complexification ~$\cg^\bac \cong \cg'^\bac$.
We call them ~{\it parabolically related}~ if they have parabolic subalgebras
~$\cp = \cm \oplus \ca \oplus \cn$,
~$\cp' = \cm' \oplus \ca' \oplus \cn'$,
such that: ~$\cm^\bac ~\cong~ \cm'^\bac$~  ($\Rightarrow \cp^\bac ~\cong~ \cp'^\bac$).\dia

Certainly, there are many such parabolic relationships for any given algebra ~$\cg$.
Furthermore, two algebras ~$\cg,\cg'$~ may be parabolically related with different
parabolic subalgebras. For example, the exceptional Lie algebras $E_{6(6)}$ and $E_{6(2)}$
are considered in Section 7 (as related also to $E_{6(-14)}$)
with maximal parabolics such that $\cm^\bac \cong \cm'^\bac \cong
sl(6,\bbc)$. But these two algebras are related also by
 another pair
of maximal parabolics ~$\tilde{\cp}^\bac,\ \tilde{\cp}'^\bac$ such that
$\tilde{\cm}^\bac \cong \tilde{\cm}'^\bac \cong
sl(3,\bbc) \oplus sl(3,\bbc)\oplus sl(2,\bbc)$, cf. \cite{Dobinv}, (11.4),(11.7).

Another interesting example are the algebras  ~$so^*(2m)$~ and ~$so(p,q)$~ which have
a series of   maximal parabolics with $\cm$-factors  \cite{Dobinv},:
\eqnn{sosopq} \cm_j ~&=&~ su^*(2j) \oplus so^*(2m-4j)\ , ~~~ j \leq [\ham] \ , \\
\cm'_j ~&=&~ sl(2j,\bbr) \oplus so(p-2j,q-2j) \ , ~~~ j \leq [\haq]\leq [\hap] \ ,   \nn\eea
whose complexifications coincide for ~$p+q = 2m$~
\eqn{sosopqc} (\cm_j)^\bac ~=~ (\cm'_j)^\bac ~=~ sl(2j,\bbc) \oplus so(2m-4j,\bbc) \ ,
\quad j \leq [\haq] \leq [\ham] = [\hapq] \ . \ee

As we know only for ~$m=2n$, i.e., for ~$so^*(4n)$~ is fulfilled
relation \eqref{relkmc}, with ~$\cm ~=~ \cm_n ~=$ $= su^*(2n)$~ from
\eqref{sosopq}, (recalling that ~$\ck' \cong su(2n)$). Obviously,
$so(p,q)$ is parabolically related to ~$so^*(4n)$~ with this
$\cm$-factor
 only when ~$p=q=2n$, i.e., $\cg'=so(2n,2n)$ with ~$\cm'_n ~=~ sl(2n,\bbr)$ (which is
 outside the range of \eqref{sosopqc}).

We leave the classification of the parabolic relations between the non-compact semisimple Lie algebras
for a subsequent publication.  In the present paper we consider mainly algebras
parabolically related to conformal Lie
algebras with maximal parabolics fulfilling \eqref{relkmc}.
We summarize the relevant cases in the following table:

\bigskip
\vbox{\offinterlineskip {\bf Table} of conformal Lie
algebras (CLA) $\cg$ with $\cm$-factor fulfilling \eqref{relkmc} \\[2pt]
\indent ~~~~~and the corresponding parabolically related algebras $\cg'$\medskip \halign{\baselineskip12pt
\strut
\vrule#\hskip0.1truecm & #\hfil&
\vrule#\hskip0.1truecm & #\hfil&
\vrule#\hskip0.1truecm & #\hfil&
\vrule#\hskip0.1truecm & #\hfil&
\vrule#\hskip0.1truecm & #\hfil&
\vrule
#\cr \tablerule &&& && &&&&&\cr
&~ $\cg$ &&~ $\ck'$ &&~  $\cm $ &&~$\cg'$&& ~$\cm'$&\cr
&&&&&&&&&&\cr
&&&&&~~$\dim V$&&&&&\cr
\tablerule
&&&&&&&&&&\cr
&~$so(n,2)$ && ~$so(n)$ &&
~$so(n-1,1)$ && $~so(p,q),$&&$so(p-1,q-1)$&\cr
& ~$n\geq3$&&&&&& $p+q=$ &&&\cr
&&&&&~~$n$&&$=n+2$&&&\cr \tablerule
&           &&&&&&&&&\cr

&~$su(n,n)$ && ~$su(n)\oplus su(n)$ &&~$sl(n,\bbc)_\bbr$&&
~$sl(2n,\bbr)$&& ~$sl(n,\bbr)\oplus sl(n,\bbr)$ &\cr & ~$n\geq
3$&&&&  &&&&&\cr &&&&&~~$n^2$ && ~$su^*(2n)$,
$n=2k$&&~$su^*(2k)\oplus su^*(2k)$&\cr \tablerule

&&&&&&&&&&\cr

&~$sp(n,\bbr)$ && ~$su(n)$ && ~$sl(n,\bbr)$ && ~$sp(r,r)$, $n=2r$
&&~$su^*(2r)$, $n=2r$&\cr &~$\rank =n\geq 3$ &&&&&&&&&\cr &&&&&
~$n(n+1)/2$&&&&&\cr \tablerule &           &&&&&&&&&\cr

&~$so^*(4n)$ && ~$su(2n)$ && ~$su^*(2n)$
&& ~$so(2n,2n)$
&&~$sl(2n,\bbr)$&\cr
&~$n\geq 3$
&&&&&&&&&\cr &&&&& ~$n(2n-1)$&&&&&\cr \tablerule
&           &&&&&&&&&\cr

&~$E_{7(-25)}$   && ~$e_6$ && ~$E_{6(-26)}$ && ~$E_{7(7)}$
&&~$E_{6(6)}$&\cr &&&&&&&&&&\cr
&&&&&~~$27$&&&&&\cr \tablerule
&           &&&&&&&&&\cr
&below not CLA ! &&&&&&&&&\cr
&           &&&&&&&&&\cr
\tablerule
&           &&&&&&&&&\cr
 &~$E_{6(-14)}$   && ~$so(10)$ && ~$su(5,1)$ && ~$E_{6(6)}$
&&~$sl(6,\bbr)$&\cr
&           &&&&&&&&&\cr
&&&&&~~$21$&&~$E_{6(2)}$&&~$su(3,3)$&\cr
&           &&&&&&&&&\cr
\tablerule
}}
\nt where we have included also the algebra ~$E_{6(-14)}\,$; we
display only the semisimple
part ~$\ck'$~ of ~$\ck$; ~$sl(n,\bbc)_\bbr$~ denotes $sl(n,\bbc)$ as
a real Lie algebra, (thus, ~$(sl(n,\bbc)_\bbr)^\bac =
sl(n,\bbc)\oplus sl(n,\bbc)$); ~$e_6$~ denotes the compact real form
of ~$E_6\,$; and we have imposed
restrictions to avoid coincidences or degeneracies due to well known isomorphisms:
~$so(1,2) \cong sp(1,\bbr) \cong su(1,1) $, ~~$so(2,2) \cong so(1,2)\oplus so(1,2)$,
~$su(2,2) \cong so(4,2)$, ~$sp(2,\bbr) \cong so(3,2)$,
  ~$so^*(4)\cong so(3) \oplus so(2,1)$, ~$so^*(8)\cong so(6,2)$.

\np

After this extended introduction we give the outline of the paper.
In Section 2 we give the
preliminaries, actually recalling and adapting facts from
\cite{Dobinv}. We add a remark on ~{\it conservation laws}~
and ~{\it conserved currents}~ which are an integral part
of our approach.
In Section 3 we consider   the case of the pseudo-orthogonal algebras
$so(p,q)$ which are parabolically related to the conformal algebra $so(n,2)$
for $p+q=n+2$. We add historical remarks and a remark on shadow representations.
In Section 4 we consider the algebras $su^*(4k)$ and $sl(4k,\bbr)$
as parabolically related to the CLA $su(2k,2k)$.
In Section 5 we consider the algebra $sp(r,r)$ as parabolically related to the CLA
$sp(2r)$ (of rank $2r$). In Section 6 we consider the algebra $E_{7(7)}$
as parabolically related to the CLA $E_{7(-25)}\,$.
In Section 7 we consider the algebras $E_{6(6)}$ and $E_{6(2)}$
as parabolically related to the hermitian symmetric case $E_{6(-14)}\,$.
In Section 8 we give Summary and Outlook.

\section{Preliminaries}

 Let $G$ be a semisimple non-compact Lie group, and $K$ a
maximal compact subgroup of $G$. Then we have an {\it Iwasawa
decomposition} ~$G=KA_0N_0$, where ~$A_0$~ is Abelian simply
connected vector subgroup of ~$G$, ~$N_0$~ is a nilpotent simply
connected subgroup of ~$G$~ preserved by the action of ~$A_0$.
Further, let $M_0$ be the centralizer of $A_0$ in $K$. Then the
subgroup ~$P_0 ~=~ M_0 A_0 N_0$~ is a {\it minimal parabolic subgroup} of
$G$.  A {\it parabolic subgroup} ~$P ~=~ M' A' N'$~ is any subgroup of $G$
which contains a minimal parabolic subgroup.

Further, let ~$\cg,\ck,\cp,\cm,\ca,\cn$~ denote the Lie algebras of ~$G,K,P,M,A,N$, resp.

For our purposes we need to restrict to  ~{\it maximal ~ parabolic
subgroups} ~$P=MAN$, i.e.  $\rank A =1$, resp. to ~{\it maximal ~ parabolic
subalgebras} ~$\cp = \cm \oplus \ca \oplus \cn$~ with ~$\dim\, \ca=1$.

Let ~$\nu$~ be a (non-unitary) character of ~$A$, ~$\nu\in\ca^*$,
parameterized by a real number ~{\it $d$}, called the {\it conformal weight} or
energy.

Further, let ~ $\mu$ ~ fix a discrete series representation
~$D^\mu$~ of $M$ on the Hilbert space ~$V_\mu\,$, or   the
finite-dimensional (non-unitary) representation of $M$ with the same
Casimirs.

 We call the induced
representation ~$\chi =$ Ind$^G_{P}(\mu\otimes\nu \otimes 1)$~ an
~{\it \it elementary representation} of $G$ \cite{DMPPT}. (These are
called {\it generalized principal series representations} (or {\it
limits thereof}) in \cite{Knapp}.)  Their spaces of functions are:  \eqnn{func}
\cc_\chi ~&=&~ \{ \cf \in C^\infty(G,V_\mu) ~ \vr ~ \cf (gman) ~=~ \\
&=& ~
e^{-\nu(H)} \cdot D^\mu(m^{-1})\, \cf (g) \} \nn\eea  where ~$a=
\exp(H)\in A'$, ~$H\in\ca'\,$, ~$m\in M'$, ~$n\in N'$. The
representation action is the {\it left regular action}:  \eqn{lrega}
(\ct^\chi(g)\cf) (g') ~=~ \cf (g^{-1}g') ~, \quad g,g'\in G\ .\ee


\bu An important ingredient in our considerations are the ~{\it \it
highest/lowest weight representations}~ of ~$\cg^\bac$. These can be
realized as (factor-modules of) Verma modules ~$V^\L$~ over
~$\cg^\bac$, where ~$\L\in (\ch^\bac)^*$, ~$\ch^\bac$ is a Cartan
subalgebra of ~$\cg^\bac$, weight ~$\L = \L(\chi)$~ is determined
uniquely from $\chi$ \cite{Dob}.

Actually, since our ERs may be induced from finite-dimensional
representations of ~$\cm$~   the Verma modules are
always reducible. Thus, it is more convenient to use ~{\it \it
generalized Verma modules} ~$\tV^\L$~ such that the role of the
highest/lowest weight vector $v_0$ is taken by the
(finite-dimensional) space ~$V_\mu\,v_0\,$. For the generalized
Verma modules (GVMs) the reducibility is controlled only by the
value of the conformal weight $d$. Relatedly, for the \idos{} only
the reducibility w.r.t. non-compact roots is essential.

\bu One main ingredient of our approach is as follows. We group the
(reducible) ERs with the same Casimirs in sets called ~{\it
multiplets}. The multiplet corresponding to fixed values of the
Casimirs may be depicted as a connected graph, the {\it vertices} of which
correspond to the reducible ERs and the {\it lines (arrows)}  between the vertices
correspond to intertwining operators.  The explicit parametrization
of the multiplets and of their ERs is important for understanding of
the situation. The notion of multiplets was introduced in \cite{Dobmul},\cite{Dobc}
and applied to representations of ~$SO_o(p,q)$~ and ~$SU(2,2)$, resp.,
induced from their minimal parabolic subalgebras. Then it was applied to
the conformal superalgebra \cite{DoPemul}, to
infinite-dimensional (super-)algebras \cite{Dobmuinf},
to quantum groups \cite{Dobqg}.\footnote{For other applications
we refer to \cite{Dobmuvar}.}

\nt {\it Remark:}~ Note that the multiplets of Verma modules include
in general more members, since there enter Verma modules which are
induced from infinite-dimensional representations of ~$\cm$~  but
nevertheless have the same Casimirs. The main multiplets in this
case contain as many members as the Weyl group $W (\cg^\bac)$ of
$\cg^\bac$. For example, for ~$su(2,2)$~ the maximal multiplets
contain 24 members ($\vr W(sl(\ell,\bbc))\vr=\ell!$), which were
considered in \cite{Dobc} and the ~$su(2,2)$~ sextets of ERs induced
from the maximal parabolic with $\cm=sl(2,\bbc)$ are submerged in
the 24-member multiplets.\dia

In fact, the multiplets contain explicitly all the data necessary to
construct the \idos{}. Actually, the data for each \ido{} consists
of the pair ~$(\b,m)$, where $\b$ is a (non-compact) positive root
of ~$\cg^\bac$, ~$m\in\bbn$, such that the {\it BGG  Verma module
reducibility condition} (for highest weight modules) is fulfilled:
\eqn{bggr} (\L+\r, \b^\vee ) ~=~ m \ , \quad \b^\vee \equiv 2 \b
/(\b,\b) \ \ee where $\r$ is half the sum of the positive roots of
~$\cg^\bac$. When the above holds then the Verma module with shifted
weight ~$V^{\L-m\b}$ (or ~$\tV^{\L-m\b}$ ~ for GVM and $\b$
non-compact) is embedded in the Verma module ~$V^{\L}$ (or
~$\tV^{\L}$). This embedding is realized by a singular vector
~$v_s$~  expressed by a polynomial ~$\cp_{m,\b}(\cg^-)$~ in the universal
enveloping algebra ~$(U(\cg_-))\ v_0\,$, ~$\cg^-$~ is the subalgebra
of ~$\cg^\bac$ generated by the negative root generators \cite{Dix}.
 More explicitly, \cite{Dob}, ~$v^s_{m,\b} = \cp_{m,\b}\, v_0$ (or ~$v^s_{m,\b} =
 \cp_{m,\b}\, V_\mu\,v_0$ for GVMs).\footnote{For
explicit expressions for singular vectors we refer to \cite{Dobsin}.}

   Then there exists \cite{Dob} an ~{\it \ido{}}~ of order ~$m=m_\b$~:
\eqn{invop}
  \cd_{m,\b} ~:~ \cc_{\chi(\L)}
~\llr ~ \cc_{\chi(\L-m\b)} \ee given explicitly by: \eqn{singvv}
 \cd_{m,\b} ~=~ \cp_{m,\b}(\widehat{\cg^-})  \ee where
~$\widehat{\cg^-}$~ denotes the {\it right action} on the functions
~$\cf\,$.

Thus, in each such situation we have an ~{\it invariant differential equation}~ of order ~$m=m_\b$~:
\eqn{invde} \cd_{m,\b}\ f ~=~ f' \ , \qquad f \in \cc_{\chi(\L)} \ , \quad
f' \in \cc_{\chi(\L-m\b)} \ .\ee

In most of these situations the invariant operator ~$\cd_{m,\b}$~ has a non-trivial invariant
kernel in which a subrepresentation of $\cg$ is realized. Thus, studying the equations
with trivial RHS:
\eqn{invdec} \cd_{m,\b}\ f ~=~ 0 \ , \qquad f \in \cc_{\chi(\L)} \ ,
   \ee
is also very important. For example, in many physical applications
 in the case of first order differential operators,
i.e., for ~$m=m_\b = 1$, equations \eqref{invdec}
are called ~{\it conservation laws}, and the elements ~$f\in \ker \cd_{m,\b}$~
are called ~{\it conserved currents}.

\md

The above construction works also for the ~{\it subsingular
vectors}~ $v_{ssv}$~ of Verma modules. Such a vector is also expressed by
a polynomial ~$\cp_{ssv}(\cg^-)$~ in the universal
enveloping algebra:
   ~$v^s_{ssv} = \cp_{ssv}(\cg^-)\, v_0\,$, cf. \cite{Dobcond}.
Thus, there exists a ~{\it conditionally invariant differential operator}
~ given explicitly by:
~$\cd_{ssv} ~=~ \cp_{ssv}(\widehat{\cg^-})$,
and a ~{\it conditionally invariant differential equation},
for many more details, see \cite{Dobcond}.
(Note that these operators/equations are not of first order.)

\md

Below in our exposition we shall use the so-called Dynkin labels: \eqn{dynk} m_i
~\equiv~ (\L+\r,\a^\vee_i)  \ , \quad i=1,\ldots,n,\ee where ~$\L =
\L(\chi)$, ~$\r$ is half the sum of the positive roots of
~$\cg^\bac$.

We shall use also   the so-called Harish-Chandra parameters:
\eqn{dynhc} m_\b \equiv (\L+\r, \b )\ ,\ee where $\b$ is any
positive root of $\cg^\bac$. These parameters are redundant, since
they are expressed in terms of the Dynkin labels, however,   some
statements are best formulated in their terms.\footnote{Clearly, both the Dynkin labels and
Harish-Chandra parameters have their origin in the BGG reducibility condition \eqref{bggr}.}

\section{The pseudo-orthogonal algebras  ~{\boldmath $so(p,q)$}}

\subsection{Choice of parabolic subalgebra}

\nt Let ~$\cg=so(p,q)$, ~$p\geq q$, ~$p+q>4$.\footnote{We shall explain
the last restriction at the end of this section.} Most of the results here
are known for ~$q=1,2$, cf. \cite{DoPe:78},\cite{PeSo},\cite{Dobsrni},\cite{Dobpeds}, and the purpose
of the consideration is to extend those for arbitrary $q$.

 For fixed ~$p,q$~ this algebra has at least ~$q$~ maximal parabolic subalgebras \cite{Dobinv}.
For example, when ~$p>q$~ there are the following possibilities for
~$\cm$-factor (cf. (7.11) of \cite{Dobinv}):
\eqn{cmsosmax} \cm^{\rm max}_j ~=~ sl(j,\bbr) \oplus so(p-j,q-j) \ , \quad j=1,2,\ldots,q\ . \ee
(There are more choices when $p=q$.)

We would like to consider a case, which would relate parabolically all $\cg=so(p,q)$ for
~$p+q$~-fixed. Thus, in order
in order to include the case $q=1$ (where there is only one parabolic which is both
minimal and maximal), we choose the case $j=1$ :
\eqn{cmsosmay} \cm ~=~ \cm^{\rm max}_1 ~=~ so(p-1,q-1) \  . \ee
Then we have:
\eqn{soprn} \dim\, \cn ~=~ \dim\, \tcn ~=~ p+q -2 \ .\ee
With this choice we get for the conformal algebra exactly
the Bruhat decomposition in \eqref{bruc}.

 We label   the signature of the ERs of
$\cg$   as follows: \eqnn{sgnd}  &&\chi ~~=~~ \{\, n_1\,, \ldots,\,
n_{\hh}\,;\, c\, \} \ , \\  &&\quad n_j \in \bbz/2\ , \quad
c=d-\han\ , \quad \hh \equiv [\han] ,\nn\\  && \vr n_1 \vr < n_2 <
\cdots <  n_{\hh}\ , \quad p+q ~~{\rm even}\ ,\nn\\  && 0 < n_1 < n_2 <
\cdots <  n_{\hh} \ , \quad p+q ~~{\rm odd}\ ,\nn\eea where the last
entry of ~$\chi$~ labels the characters of $\ca\,$, and the first
$\hh$ entries are labels of the finite-dimensional nonunitary irreps
of $\cm\cong so(p-1,q-1) $.

The reason to use the parameter ~$c$~ instead of ~$d$~ will become clear below.

\subsection{Main multiplets}

Following results of \cite{DoPe:78,PeSo,Dobsrni},\cite{Dobpeds} we present the main multiplets
(which contain the maximal number of ERs with this parabolic)
with the explicit parametrization of the ERs in the multiplets  in a simple
  way (helped by the use of the signature entry ~$c$):
\eqnn{sgne}  \chi^\pm_1 &=& \{ \eps\,  n_1\,, \ldots,\,
n_\hh \,;\, \pm n_{\hh+1} \} \ ,\\ && \quad n_\hh < n_{\hh+1}\ , \nn\\
\chi^\pm_2 &=& \{ \eps\,  n_1\,, \ldots,\, n_{\hh-1}\,,\,
n_{\hh+1}\,;\, \pm n_\hh \}     \nn\\  \chi^\pm_3  &=&  \{  \eps\,
n_1,  \ldots,  n_{\hh-2},  n_{\hh},  n_{\hh+1}\,;\, \pm
n_{\hh-1} \}     \nn\\  ... \nn\\
 \chi^\pm_{\hh-1} &=& \{ \eps\, n_1\,,
n_2\,, n_4\,,\ldots,\, n_{\hh}\,,\, n_{\hh+1}\,;\, \pm  n_3 \}  \nn\\
 \chi^\pm_{\hh} &=& \{ \eps\, n_1\,,
n_3\,, \ldots,\, n_{\hh}\,,\, n_{\hh+1}\,;\, \pm  n_2 \}     \nn\\
\chi^\pm_{\hh+1} &=& \{ \eps\, n_2\,, n_3\,, \ldots,\, n_{\hh}\,,\,
n_{\hh+1}\,;\, \pm  n_1 \}     \nn\\  &\eps =& \begin{cases}
 \pm\,, ~&~  p+q ~~ even  \nn\\
                     1,  ~&~  p+q  ~~ odd \end{cases} \nn\eea
($\eps = \pm$~ is correlated with $\chi^\pm$).
Clearly, the multiplets correspond 1-to-1 to the finite-dimensional irreps
of $so(p+q,\bbc)$ with signature $\{n_1,\ldots,n_h,n_{h+1}\}$ and we are able to use
previous results due to the parabolic relation between the $so(p,q)$ algebras
for $p+q$~-fixed.

Note that the two representations in each pair ~$\chi^\pm$~ were called ~{\it shadow fields}~
in the 1970s, see more on this  towards the end of this Section.

Further, the number of ERs in the corresponding multiplets is equal to ~$2[\hal]=2(h+1)$.
This maximal number is equal to the following ratio of numbers of elements of Weyl groups:
\eqn{multi} \vr W(\cg^\bac,\ch^\bac)\vr\, /\, \vr
W(\cm^\bac,\ch_m^\bac)\vr \ , \ee
where ~$\ch^\bac,\ \ch^\bac_m$~ are Cartan subalgebras of ~$\cg^\bac,\ \cm^\bac$, resp.

The above formula actually holds for all conformal Lie algebras
 and those parabolically related to them. More precisely, we have:

\nt\bu  {\it The number of elements of the main multiplets   of a
conformal Lie algebra $\cg$ with $\cm$-factor fulfilling \eqref{relkmc}
is given by \eqref{multi}. The same number holds for
any algebra $\cg'$ parabolically related to $\cg$ w.r.t. $\cm$.}\dia

 Further, we denote by ~$\cc^\pm_i$~ the representation space with signature
~$\chi^\pm_i\,$.

We first give the multiplets pictorially in Figures 1 and 2 for $p+q$
even and odd, resp., and then explain notations and results:

\np\thispagestyle{empty}

\fig{}{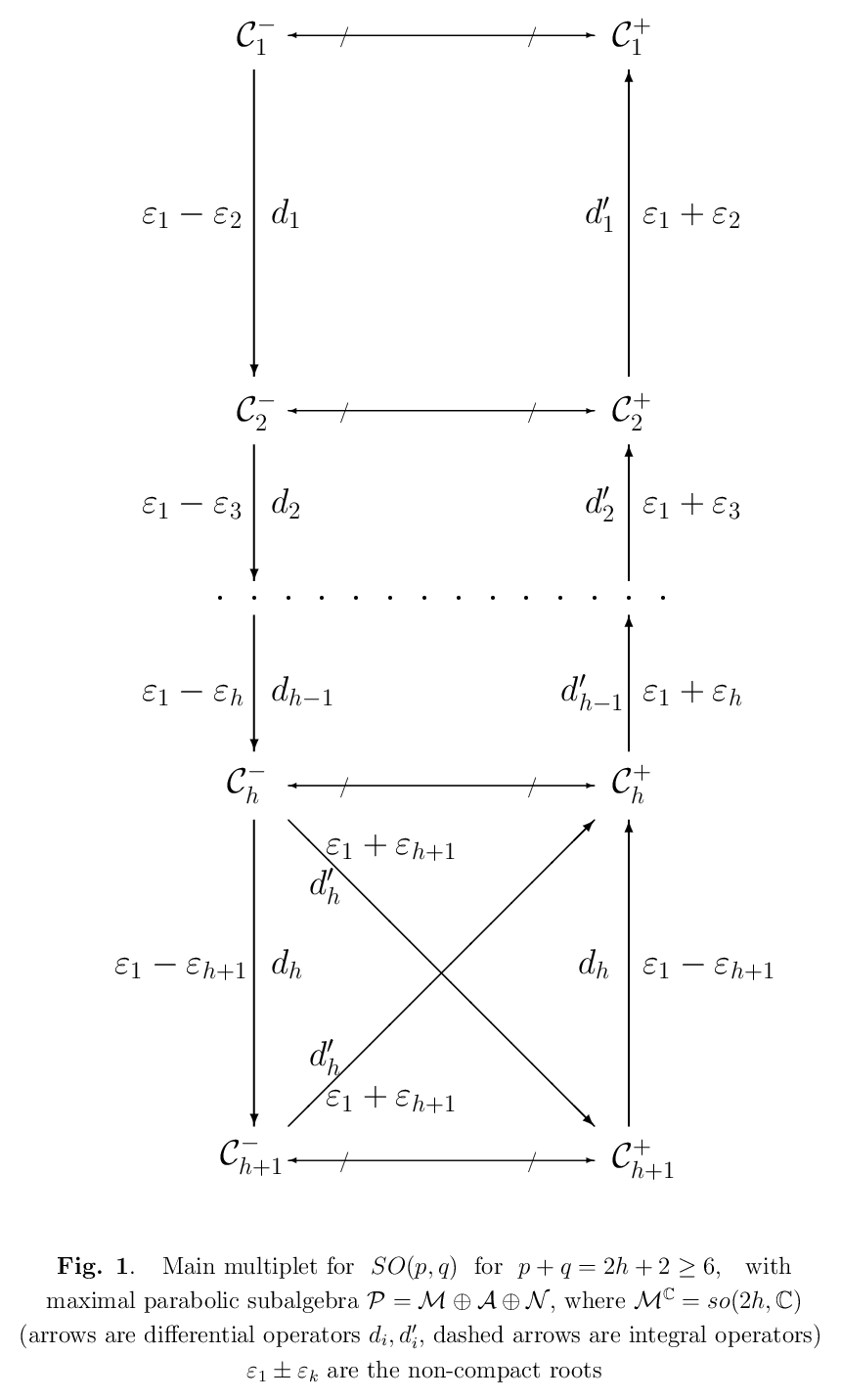}{14cm}

\thispagestyle{empty} \fig{}{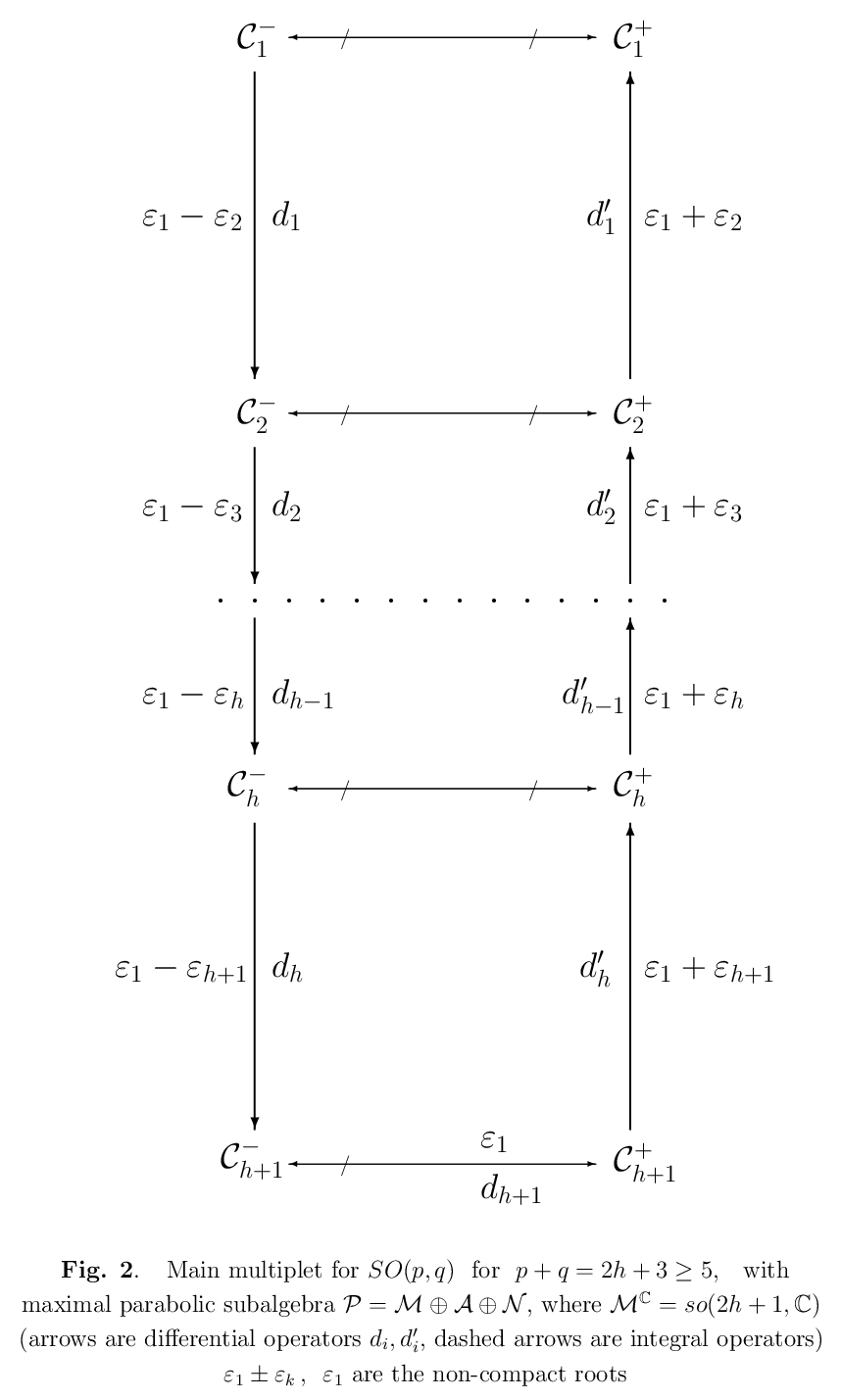}{14cm}


The ERs in the multiplet are related by {\it intertwining integral and
differential operators}.

The  {\it integral operators} were introduced by
Knapp and Stein \cite{KnSt}. They correspond to elements of the restricted Weyl group of $\cg$.
In fact, these operators are defined for any ER,
not only for the reducible ones, the general action being in the context of
\eqref{sgnd},\eqref{sgne}~:
\eqnn{knast}  G  ~&:&~ \cc_\chi ~ \llr ~ \cc_{\chi'} \ ,\\
\chi  ~&=&~  \{\, n_1\,, \ldots,\,
n_{\hh}\,;\, c\, \} \ , \nn\\
\chi' ~&=&~ \{\, (-1)^{p+q+1}n_1\,, \ldots,\, n_{\hh}\,;\, -c\, \} \ . \nn\eea

These operators intertwine the pairs ~$\cc^\pm_i$~ (cf. \eqref{sgne}):
\eqn{knapps}
G^\pm_i ~:~ \cc^\mp_i \lra \cc^\pm_{i}   \ , \quad i ~=~
1,\ldots,1+\hh  \ . \ee

In the conformal setting (both Euclidean $q=1$ and Minkowskian
$q=2$) the integral kernel of the Knapp-Stein operator is given by
the conformal two-point function \cite{DMPPT}.

\md

The {\it \idos}\ correspond to non-com\-pact positive roots of the root
system of ~$so(p+q,\bbc)$, cf. \cite{Dob}. In the current context, compact
roots of $so(p+q,\bbc)$ are those that are roots also of the
subalgebra $so(p+q-2,\bbc)$, the rest of the roots are non-compact.
In more detail, we briefly recall the root systems:

For ~$p+q=2h+2$ ~even,   the positive root system of
~$so(2h+2,\bbc)$~ may be given by vectors ~$\veps_i \pm \veps_j\,$,
~$1\leq i <j \leq h+1$, where ~$\veps_i$~ form an orthonormal basis
in $\bbr^{h+1}$, i.e., ~$(\veps_i,\veps_j) = \d_{ij}\,$. The
non-compact roots may be taken as ~$\veps_{1} \pm \veps_i\,$,
~$2\leq i \leq h+1$. The root ~$\veps_{1} - \veps_i\,$
corresponds to the operator ~$d_{i-1}\,$, the root
~$\veps_{1} + \veps_i\,$  corresponds to the
operator ~$d'_{i-1}\,$.

For  ~$p+q=2h+3$ ~odd, the positive root system of
~$so(2h+3,\bbc)$~ may be given by vectors ~$\veps_i \pm
\veps_j\,$, ~$1\leq i <j \leq h+1$, ~$\veps_k$, ~$1 \leq k \leq
h+1$. The non-compact roots may be taken as ~$\veps_{1} \pm
\veps_i\,$, ~$\veps_{1}\,$.
The root ~$\veps_{1} - \veps_i\,$
  corresponds to the operator ~$d_{i-1}\,$,
the root ~$\veps_{1} + \veps_i\,$
corresponds to the operator ~$d'_{i-1}\,$.
The root ~$\veps_{1}$~ has a special position since
it intertwines the same ERs that are intertwined by the
Knapp-Stein integral operator ~$G^+_{h+1}\,$. The latter means
that ~$G^+_{h+1}\,$ degenerates to the differential operator ~$d_{h+1}\,$,
and this degenerations determines that ~$d_{h+1}~\sim ~\Box^{\,n_1}\,$,
(for $n_1\in\bbn$), where ~$\square$~ is the d'Alembert operator,
as explained explicitly for the case ~$so(3,2)$~ in  \cite{Dobso}.
   (The latter phenomenon
happens   for the Knapp-Stein integral operators at
 critical points, but usually there is no
non-compact root involved, cf., e.g.,  \cite{DMPPT}.)

The degrees of these \idos\ are given just by the differences of the
~$c$~ entries \cite{Dobsrni}: \eqnn{degr}&& \deg d_i = \deg
d'_i = n_{\hh+2-i} - n_{\hh+1-i} \,, \qquad i = 1,\ldots,h \,,
  \\  &&\deg d'_{h} = n_2 + n_1
\,, \quad p+q ~~ {\rm even}\ , \nn\\  &&\deg d_{h+1} = 2n_1
\,, \quad p+q ~~ {\rm odd} \ .
\nn\eea
where $d'_h$ is omitted from the first line for $(p+q)$ even.
By our construction they are equal to
the  Harish-Chandra parameters for the non-compact roots:
\eqnn{deghc}&& \deg d_i ~=~ m_{\veps_1-\veps_{i+1}}  \ , \\
&& \deg d'_i ~=~  m_{\veps_1+\veps_{i+1}} \ , \qquad i = 1,\ldots,h \,, \nn\\
&& \deg d_{h+1} ~=~ m_{\veps_1} \ .
   \eea

Matters are arranged so that in every multiplet only the ER with
signature ~$\chi^-_1$~ contains a {\it finite-dimensional nonunitary
subrepresentation} in  a   subspace ~$\ce$. The
latter corresponds to the finite-dimensional unitary irrep of
~$so(p+q)$~ with signature  ~$\{ n_1\,, \ldots,\, n_\hh \,, \,
n_{\hh+1} \}$. The subspace ~$\ce$~ is annihilated by the operator
~$G^+_1\,$,\ and is the image of the operator ~$G^-_1\,$.

Although the diagrams are valid for arbitrary  $so(p,q)$ ($p+q\geq 5$) the contents
is very different. We comment only on the ER with signature
~$\chi^+_1\,$. In all cases it contains  an UIR of $so(p,q)$ realized on an
invariant subspace ~$\cd$~ of
the ER ~$\chi^+_1\,$. That subspace is annihilated by the operator
~$G^-_1\,$,\ and is the image of the operator ~$G^+_1\,$.
(Other ERs contain more UIRs.)

If ~$pq \in 2\bbn$~ the mentioned UIR is a discrete series representation.
Other ERs contain more discrete series UIRs. The number of discrete series is given
by the formula \cite{Knapp}:
\eqn{wwgwwh} \vr
W(\cg^\bac,\ch^\bac)\vr\, /\, \vr W(\ck^\bac,\ch^\bac)\vr \ , \ee
where ~$\ch^\bac$~ is a Cartan subalgebra of both ~$\cg^\bac$~ and
~$\ck^\bac$.

And if ~$q=2$~ the invariant subspace ~$\cd$~ is the direct sum of two subspaces
~$\cd ~=~ \cd^+ \oplus \cd^-$, in which are realized a
 {\it holomorphic discrete series representation} and its conjugate
  {\it   anti-holomorphic discrete
series representation}, resp. These are contained only in ~$\chi^+_1$~ and
count for two series in the formula \eqref{wwgwwh}.
Furthermore, any  holomorphic discrete series representation is infinitesimally equivalent to a
{\it lowest weight GVM} of the conformal algebra $so(p,2)$, while
an anti-holomorphic discrete series representation is infinitesimally equivalent to a
{\it highest weight GVM}.\\
Highest/lowest weight GVMs are related to other pairs besides ~$\chi^+_1$. \\
A detailed analysis of these  occurrences is done for the conformal algebra
$so(3,2)$ in \cite{Dobpeds} and for $so(4,2)$ in \cite{PeSo},\cite{Dobpeds}.

\subsection{Reduced multiplets}

Besides the main multiplets which are 1-to-1 with the finite-dimensional irreps of
$so(p+q,\bbc)$,   there are other multiplets which we describe here.

\bu We start with the case ~$p+q$~even. In this case there are ~$h+1$($=(p+q)/2$)~
multiplets - doublets - each consisting of a pair with signatures ~$\tchi^\pm$~ given
explicitly as follows:
\eqnn{sgner}  \tchi^\pm_1 &=& \{ \pm\,  n_1\,, \ldots,\,
n_\hh \,;\, \pm n_{\hh} \}  \\
\tchi^\pm_2 &=& \{ \pm\,  n_1\,, \ldots,\, n_{\hh-1}\,,\,
n_{\hh+1}\,;\, \pm n_{\hh-1} \}     \nn\\  \tchi^\pm_3  &=&  \{  \pm\,
n_1,  \ldots,  n_{\hh-2},  n_{\hh},  n_{\hh+1}\,;\, \pm
n_{\hh-2} \}     \nn\\  ... \nn\\
 \chi^\pm_{\hh-1} &=& \{ \pm\, n_1\,,
n_2\,, n_4\,,\ldots,\, n_{\hh}\,,\, n_{\hh+1}\,;\, \pm  n_2 \}   \nn\\
 \tchi^\pm_{\hh} &=& \{ \pm\, n_1\,,
n_3\,, \ldots,\, n_{\hh}\,,\, n_{\hh+1}\,;\, \pm  n_1 \} \,, ~~~n_1 \neq 0     \nn\\
\tchi^\pm_{\hh+1} &=& \{ \mp\, n_1\,, n_3\,, \ldots,\, n_{\hh}\,,\,
n_{\hh+1}\,;\,  \pm n_1   \}\,, ~~~n_1 \neq 0 \nn \eea
Clearly, the  signature ~$\tchi^\pm_i$~ may be obtained from ~$\chi^\pm_i$~
by setting the corresponding Harish-Chandra parameter equal to zero:
\eqnn{reduc} &&m_{\veps_1\pm \veps_{i+1}} = \deg d_i = \deg
d'_i = n_{\hh+2-i} - n_{\hh+1-i} ~=~ 0 \,, \qquad i = 1,\ldots,h-1 \,, \nn\\
&&m_{\veps_1 - \veps_{h+1}} = \deg d_h =  n_2 - n_1  ~=~ 0 \,, ~~~{\rm for}~\tchi^\pm_{\hh} \ ,\\
&&m_{\veps_1 + \veps_{h+1}} = \deg d'_h =  n_2 + n_1  ~=~ 0 \,, ~~~{\rm for}~\tchi^\pm_{\hh+1} \ .\eea

Although written compactly as \eqref{sgne} no  pair is  related to any other pair. This may be seen
easily as follows. Consider ~$\eqref{sgne}$~ and set formally ~$n_{h+1}=n_h\,$. The signatures
~$\chi^\pm_1$~ and ~ $\chi^\pm_2$~ coincide are become equal to ~$\tchi^\pm_1\,$, but the rest of
the signatures ~$\chi^\pm_i\,$, ~$i\geq 3$~ are not allowed in our class, e.g.,
$$ \chi^\pm_3  ~\lra  \{  \eps\,
n_1,  \ldots,  n_{\hh-2},  n_{\hh},  n_{\hh}\,;\, \pm
n_{\hh-1} \} $$
is not allowed since it violates \eqref{sgnd}
due to equality of two  $\cm$-signature entries ($n_{\hh}$).
Thus, from the whole multiplet only the pair ~$\tchi^\pm_1\,$ remains in our class.
Similarly for the rest of the pairs.

Inside a fixed  pair ~$\tchi^\pm_i\,$, ~$i=1,\ldots,h+1$,
act  two operators:  a Knapp-Stein integral operator from ~$\tchi^+_i\,$ to
~$\tchi^-_i\,$,
  and a differential operator   from ~$\tchi^-_i\,$ to
~$\tchi^+_i\,$. In more detail:\\
\bu Let first ~$i=1,\ldots,h-1$.
Inside a fixed  pair ~$\tchi^\pm_i\,$,
acts the Knapp-Stein integral operator
~$G^-_i$~ \eqref{knapps} (coinciding with ~$G^-_{i+1}$~ for this signature),
and a differential operator ~$\tilde{d_i}$~ of degree ~$2n_{h+1-i}$~ which is a degeneration of the
Knapp-Stein integral operator ~$G^+_i$~
(coinciding with ~$G^+_{i+1}$~ for this signature). For this
differential operator for $n_1=0$~ we have:
~$\tilde{d_i} ~\sim~ \Box^{\,n_{h+1-i}}\,$, ($n_{h+1-i}\in\bbn$).\footnote{For $so(4,2)$, ($h=2,i=1$),
when ~$n_1=0,n_2=1$~ the latter d'Alembert operator arises also as a conditionally
invariant differential operator due to the presence of a subsingular
vector in the corresponding Verma module  \cite{Dobcond}.}
\\
\bu Inside the fixed  pair ~$\tchi^\pm_h\,$\
acts the Knapp-Stein integral operator
~$G^-_h$~ \eqref{knapps} (coinciding with ~$G^-_{h+1}$~ for this signature),
and the differential operator ~$d'_h$~ of degree ~$2n_1$~ (cf. the previous subsection)
which in addition is a degeneration of the
Knapp-Stein integral operator ~$G^+_h$~
(coinciding with ~$G^+_{h+1}$~ for this signature).\\
\bu Inside the fixed  pair ~$\tchi^\pm_{h+1}\,$\
acts the Knapp-Stein integral operator
~$G^-_{h+1}$~ \eqref{knapps} (coinciding with ~$G^+_{h}$~ for this signature),
and the differential operator ~$d_h$~ of degree ~$2n_2$~ which in addition is a degeneration of the
Knapp-Stein integral operator ~$G^+_{h+1}$~
(coinciding with ~$G^-_{h}$~ for this signature).

\md

\bu We continue with the case ~$p+q$~odd. In this case there are ~$h$~ doublets\footnote{In the
case $so(3,2)$ there are two additional doublets \cite{Dobpeds} involving the two singleton
representations, which are special for $so(3,2)$.}
  with signatures ~$\hchi^\pm$~   given
similarly to the even case as follows:
\eqnn{sgnor}  \hchi^\pm_1 &=& \{    n_1\,, \ldots,\,
n_\hh \,;\, \pm n_{\hh} \}  \\
\hchi^\pm_2 &=& \{    n_1\,, \ldots,\, n_{\hh-1}\,,\,
n_{\hh+1}\,;\, \pm n_{\hh-1} \}     \nn\\  \hchi^\pm_3  &=&  \{
n_1,  \ldots,  n_{\hh-2},  n_{\hh},  n_{\hh+1}\,;\, \pm
n_{\hh-2} \}     \nn\\  ... \nn\\  \hchi^\pm_{\hh} &=& \{   n_1\,,
n_3\,, \ldots,\, n_{\hh}\,,\, n_{\hh+1}\,;\, \pm  n_1 \}
   \nn \eea
The  signature ~$\hchi^\pm_i$~ may be obtained from ~$\chi^\pm_i$~
by setting the corresponding Harish-Chandra parameter equal to zero:
\eqn{reduco} m_{\veps_1\pm \veps_{i+1}} = \deg d_i = \deg
d'_i = n_{\hh+2-i} - n_{\hh+1-i} ~=~ 0 \,, \qquad i = 1,\ldots,h \ .\ee

Inside a fixed  pair ~$\hchi^\pm_i\,$, ~$i=1,\ldots,h$,\
acts the Knapp-Stein integral operator
~$G^-_i$~ \eqref{knapps} (coinciding with ~$G^-_{i+1}$~ for this signature),
and a differential operator ~$\hat{d_i}$~ of degree ~$2n_{h+1-i}$~which is a degeneration of the
Knapp-Stein integral operator ~$G^+_i$~
(coinciding with ~$G^+_{i+1}$~ for this signature).
For the differential operators we have ~$\hat{d_i} ~\sim~ \Box^{\,n_{h+1-i}}\,$,
(when $n_{h+1-i}\in\bbn$).
 The difference with the even situation is only for $i=h$, where the
 degeneration of ~$G^+_{h+1}$~ was present already in the main multiplet.

If ~$pq \in 2\bbn$~ the  representations ~$\tchi^+_1$, ~$\hchi^+_1$, contain an UIR
called limits of the  discrete series representations.
And if ~$q=2$~ that UIR is the
direct sum of two subspaces  in which are realized
 {\it limits of holomorphic discrete series representation} and its conjugate
  {\it limits of  anti-holomorphic discrete
series representation}, resp.  The latter do not happen in any other doublet,
while limits of   discrete series representations happen in other doublets.
(For more on this see \cite{Dobpeds} for $so(3,2)$
and \cite{PeSo},\cite{Dobpeds} for $so(4,2)$.)

\subsection{Remarks on shadow fields and history}

\bu We labelled the signature of the ERs in \eqref{sgnd} as
$$ \chi ~~=~~ \{\, n_1\,, \ldots,\, n_{\hh}\,;\, c\, \} $$
using the parameter ~$c$~ instead of the conformal weight ~$d = c +
\han$. This was used already in \cite{DMPPT} since the multiplets
were given more economically in terms of pairs of ERs in which the
parameter ~$c$~ just changes sign. (Also mathematicians  use the
parameter $c$ due to the fact that in its terms the representation
parameter space looks simple:~ the principal unitary series
representation induced from a  maximal parabolic is given by
$c=i\rho$, $\rho\in\bbr$;~ the supplementary series of unitary
representations is given by $-s < c < s$, $s\in\bbr$, etc.)

Otherwise in the current context we should use for each Knapp-Stein operators conjugated doublet of
shadow fields~:
  \eqnn{sgndd}  &&\chi^+ ~~=~~ [\, n_1\,, \ldots,\,
n_{\hh}\,;\, d \, ] \ , \qquad n_j \in \bbz/2\ , \\
&&\chi^- ~~=~~ [\,(-1)^{p+q+1} n_1\,, \ldots,\,
n_{\hh}\,;\, d_{\rm shadow}= p+q-2-d \, ] \ .
\nn\eea

The reason the representations $\chi^\pm$ in the 1970s were called "shadow fields"
in the context of the conformal algebra $so(n,2)$ is that the sum
of their conformal weights equals the dimension $n$ of Minkowski space-time - isomorphic to
$\cn$ or $\tcn$, cf. \eqref{soprn}. This continues to be true for general $so(p,q)$~:
\eqn{shad} d + d_{\rm shadow}~=~ p+q-2 ~=~ n \ , \ee
and also for all conformal Lie algebras considered in the next Sections.

Shadow fields appear all the time in conformal field theory.
For example, in \cite{Dobads} we showed that  in the generic case each   field
on the AdS bulk has ~{\it two}~ boundary fields which are shadow fields being
related by a integral Knapp-Stein operator.  Later Klebanov-Witten \cite{KleWit}
showed that these two boundary fields are related by a Legendre transform.

For a current discussion on shadow fields we refer to \cite{Mets}.

\bu The diagram for ~$p+q$~ even appeared first for the Euclidean conformal group
in four-dimensional space-time ~$SU^*(4) \cong Spin(5,1)$~ in
\cite{DoPe:78}. Later it was generalised to the Minkowskian
conformal group in four-dimensional space-time ~$SO(4,2)$~ in \cite{PeSo}.
In both cases, the three ($=(p+q)/2$) doublets (from the previous subsection)
 were also given together the corresponding degeneration of the Knapp-Stein
 integral operators.

The exposition above including  Figures 1 \& 2
follows the exposition for Euclidean case ~$so(n+1,1)$~ in \cite{Dobsrni}.
Later the results were generalised to the Minkowskian case ~$so(n,2)$ \cite{Dobpeds}.

\bu Actually, the case of Euclidean conformal group in arbitrary dimensions
~$SO(p,1)$~ was studied in \cite{DMPPT} for representations of $\cm = so(p-1)$
which are symmetric traceless tensors. This means in \eqref{sgnd} we
should set ~$n_1=n_2= \cdots = n_{h-1}=0$, and then only the first two  pairs
~$\chi^\pm_1,\chi^\pm_2$~  in \eqref{sgne}  are possible. Thus from the two figures
only the upper quadrants are relevant, and were given in \cite{DMPPT}, cf. Fig. 1 there.

\bu Above we restricted to ~$p+q\geq 5$. The excluded cases are: ~$so(3,1)$,
~$so(2,2) \cong so(2,1)\oplus so(2,1)$,
~$so(2,1)$, ($so(1,1)$ is abelian).

In the case ~$so(3,1) \cong sl(2,\bbc)$~ the multiplet in general contains only four ERs,
and is in fact representable by the diagram in the
case of symmetric traceless tensors of $so(p,1)$, $p>3$, cf.
 \cite{DMPPT}, Appendix B.

The case ~$so(2,1) \cong sl(2,\bbr)$~  is special   and must be  treated separately.
But in fact,  it is contained in what we presented already. In that
case the multiplets contain only  two ERs which may be depicted
by the  top pair $\chi^\pm_1$  in both Figures.
 (Formally, set $h=0$ in both Figures.)
They have the properties that we described, including the
 (anti)holomorphic discrete series which are present in this case.
That case was the first given already in 1946-7 independently by
Gel'fand et al \cite{GeNa} and Bargmann \cite{Barg}.


\section{The Lie algebras {\boldmath  $su^*(2n)$} and {\boldmath  $sl(n,\bbr)$}}

\subsection{Case {\boldmath  $su^*(2n)$}}

Let  ~$\cg=su^*(2n)$. It has maximal compact subalgebra ~$\ck =
sp(n)$, and thus $\cg$ does not have discrete series representations
(as $\rank\ck = n <  \rank su^*(2n)=2n-1$).

The algebra ~$\cg=su^*(2n)$~  has $n-1$ maximal parabolic
subalgebras with $\cm$-factors (cf. (5.8) from \cite{Dobinv}): \eqn{cmsmax} \cm^{\rm max}_k ~=~
su^*(2k) \oplus su^*(2(n-k)) \ ,\qquad 1\leq k \leq n-1\ , \ee with complexification:
\eqn{cmsmaxc} (\cm^{\rm max}_k)^\bac ~=~ sl(2k,\bbc) \oplus
sl(2(n-k),\bbc) \  . \ee

We would like to relate parabolically this algebra with the appropriate conformal
Lie algebra, namely, with ~$su(n,n)$. It was considered in
\cite{Dobsunn} with $\cm$-factor: ~$\cm' = sl(n,\bbc)_\bbr\,$~ which
has complexification:
\eqn{sunnmf}
\cm'^\bac = sl(n,\bbc) \oplus sl(n,\bbc) \ .\ee
Clearly, the latter expression can match \eqref{cmsmaxc} only if
~$n=2k$, i.e., ~$n$~ must be ~{\it even}.

Thus, we set ~$n=2k$~ and consider:
\eqnn{cmsust}  \cg ~&=&~su^*(4k) \ , \\
\cm ~&=&~ su^*(2k) \oplus su^*(2k) \ , \nn\\
\cm^\bac ~&=&~ sl(2k,\bbc) \oplus sl(2k,\bbc) \ .
\nn \eea

\subsection{Case {\boldmath  $sl(n,\bbr)$}}

Let  $sl(n,\bbr)$. Its maximal compact subalgebra is $\ck = so(n)$, and
thus it does not have discrete series representations. It has $[\hann]$ maximal parabolic
subalgebras obtained by deleting a node from its standard Dynkin diagram and
taking into account the symmetry (cf. \cite{Dobinv}):

\eqn{cmm} \cm_j  ~=~  sl(j,\bbr)\oplus sl(n-j,\bbr) \
, ~~~1\leq j\leq [\hann] \ . \ee
We would like to match this with both \eqref{sunnmf} and \eqref{cmsust}. Obviously this can happen
only for ~$n=4k$ and ~$j = n/2=2k$, so we consider:\footnote{If we would like to match \eqref{cmm} only
with \eqref{sunnmf} this is possible for ~$n=2k$~ and ~$j = n/2=k$.  The simplest example for this when ~$n=6$,
parabolically relating ~$sl(6,\bbr)$~ with ~$su(3,3)$, was given in \cite{Dobsunn,Dobsu33}.}
\eqnn{cmsustz}  \cg ~&=&~ sl(4k,\bbr) \ , \\
\cm  ~&=&~  sl(2k,\bbr)\oplus sl(2k,\bbr) \ , \nn\\
\cm^\bac ~&=&~ sl(2k,\bbc) \oplus sl(2k,\bbc) \ .
\nn \eea

\subsection{Representations and multiplets}

Above we have chosen the $\cm$-factors
of the Lie algebras ~$su^*(4k)$~ and ~$sl(4k,\bbr)$~ so that they are
parabolically related to the conformal Lie algebra ~$su(2k,2k)$~
with $\cm$-factor ~$\cm^\bac = sl(2k,\bbc) \oplus sl(2k,\bbc)$,
cf. \eqref{cmsust}, \eqref{cmsustz},
thus, we shall discuss them together.

The signature of the ERs of both $\cg$  may be denoted as:
\eqnn{sgnsukk}   \chi ~&=&~ \{\, n_1\,, \ldots,\, n_{2k-1}\,,\, n_{2k+1}\,
\ldots,\, n_{4k-1}\,;\, c\, \} \ ,  \\ && \quad n_j \in \bbn\ , \quad c =
d- 2k \ ,\nn\eea
same as for $su(2k,2k)$.

The   Knapp--Stein   restricted Weyl reflection mapping $\chi$ to its shadow is given by:
 \eqnn{knasta}  && G  ~:~ \cc_\chi ~ \llr ~ \cc_{\chi'} \
,\\
\chi'     &=&     \{
(n_1,\ldots,n_{2k-1},n_{2k+1},\ldots,n_{4k-1})^*  ;  -c  \} \ ,\qquad\nn  \\
&&  (n_1,\ldots,n_{2k-1},n_{2k+1},\ldots,n_{4k-1})^* ~\doteq~   \nn \\
 && (n_{2k+1},\ldots,n_{4k-1},n_1,\ldots,n_{2k-1}) \nn\eea

Further, we   use the root system of the complex algebra
~$sl(4k,\bbc)$.  The positive roots ~$\a_{ij}$~ in terms of the
simple roots ~$\a_i$~ are:
\eqnn{poskk} &&\a_{ij} = \a_i +\a_{i+1} + \cdots + \a_{j}\ ,
\qquad 1\leq i<j\leq 4k-1  \ , \\
&&\a_{ii} \equiv \a_i \ , ~~~1\leq i \leq 4k-1 \nn\eea
from which the  non-compact are:
 $$ \a_{ij}\ , \qquad 1
\leq i \leq 2k \ ,  \qquad 2k \leq j \leq 4k-1 \ $$

 The correspondence between the signatures $\chi$
and the highest weight $\L$  is through the Dynkin
labels:   \eqnn{corrsunn}  &&n_i =  m_i ~\equiv~ (\L+\r,\a^\vee_i) ~=~ (\L+\r,
\a_i )\ ,  \qquad i=1,\ldots,4k-1,  \\
      &&c = -\ha (m_\ta +
m_{2k}) =    -\,\ha(   m_1+\cdots +   m_{2k-1} + 2m_{2k} + m_{2k+1} + \cdots +
m_{4k-1})   \nn\eea   $\L = \L(\chi)$, ~~$\ta ~=~ \a_1 + \cdots   + \a_{4k-1}$~ is the
highest root.

In our diagrams we need  also the Harish-Chandra parameters for the non-compact roots
using the following notation:
$$m_{ij} \equiv m_{\a_{ij}} = m_i + \cdots + m_j\,, ~~~i<j \ $$

The number of ERs in the corresponding multiplets is according to \eqref{multi}:
\eqn{multik} \frac{\vr W(\cg^\bac,\ch^\bac)\vr}{\vr
W(\cm^\bac,\ch_m^\bac)\vr} ~=~ \frac{\vr W(sl(4k,\bbc))\vr}{\vr W(sl(2k,\bbc))\vr^2}
~=~ \frac{(4k)!}{ ((2k)!)^2} ~=~ \left( {4k\atop 2k}\right)
\ee
(which was given for $su(n,n)$ in \cite{Dobsunn}).

Below we give the diagrams for the cases ~$k=1,2$. Of course, the case ~$k=1$~
is known long time ago, first as ~$su^*(4)\cong so(5,1)$, cf. \cite{DoPe:78},
then as $su(2,2)\cong so(4,2)$, cf. \cite{PeSo}, and also as ~$sl(4,\bbr) \cong
so(3,3)$, as we recalled already
in the previous section on $so(p,q)$ algebras.  We present it here using
 a new diagram look which can handle the more complicated
cases that follow further.  In this new look only the invariant differential operators
are presented explicitly. The integral Knapp-Stein operators, more precisely the restricted Weyl
reflection action is understood by a symmetry of the picture, either w.r.t. a central point, or w.r.t.
middle line.

 Thus, in Figure 3 we give the case $k=1$, where the Knapp-Stein symmetry is
w.r.t. to the bullet in the middle of the figure. Then in   Figure 4 we
give the diagram Figure 1 for the special case  $h=2$, as given originally
for ~$so(5,1)$~ in \cite{DoPe:78}, and ~$so(4,2)$~ in \cite{PeSo},
stressing that both Figures 3 and 4 have the same content.

Next we give the case $k=2$, in Figure 5, which applies to  $su^*(8)$, $sl(8,\bbr)$ and $su(4,4)$.
(For reduced multiplets we refer to \cite{Dobsunn}.)
The diagram is very complicated and just to be able to depict all the relevant information
we must use the following condensing conventions. Each \ido\ is
represented by an arrow accompanied by a symbol ~$i_{j...\ell}$~
encoding the root ~$\b_{j...\ell}$~ and the number $m_{\b_{j...\ell}}$
which is involved in the BGG criterion. This notation is used to
save space, but it can be used due to
 the fact that only \idos\ which are
non-composite are displayed, and that the data ~$\b,m_\b\,$, which
is involved in the embedding  ~$V^\L \lra V^{\L-m_\b,\b}$~ turns out
to involve only the ~$m_i$~ corresponding to simple roots, i.e., for
each $\b,m_\b$ there exists ~$i = i(\b,m_\b,\L)\in \{ 1,\ldots,r\}$, ($r = \rank \cg$),
such that ~$m_\b=m_i\,$. Hence the data
~$\b_{j...\ell}\,$,$m_{\b_{j...\ell}}$~ is represented by ~$i_{j...\ell}$~ on
the arrows.

\fig{}{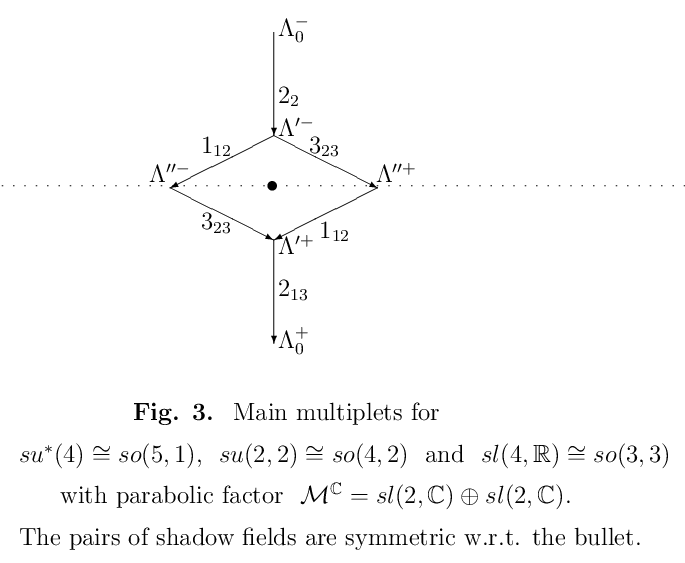}{10cm}


\fig{}{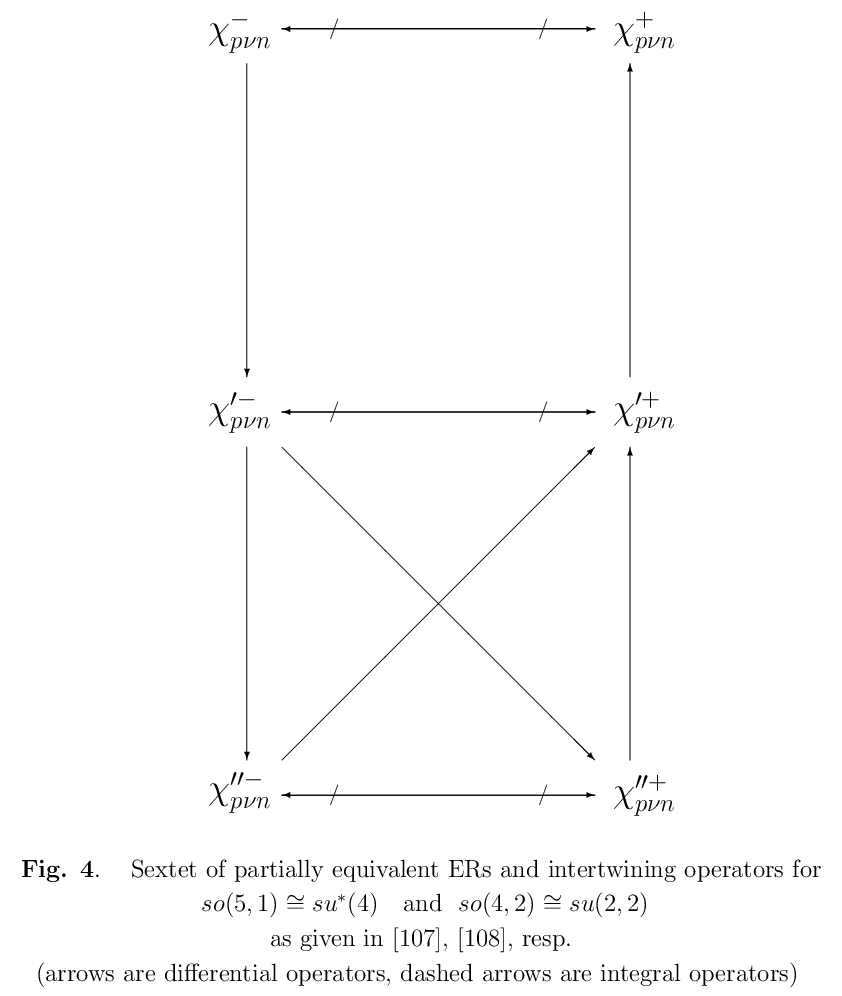}{17cm}

\thispagestyle{empty}

\fig{}{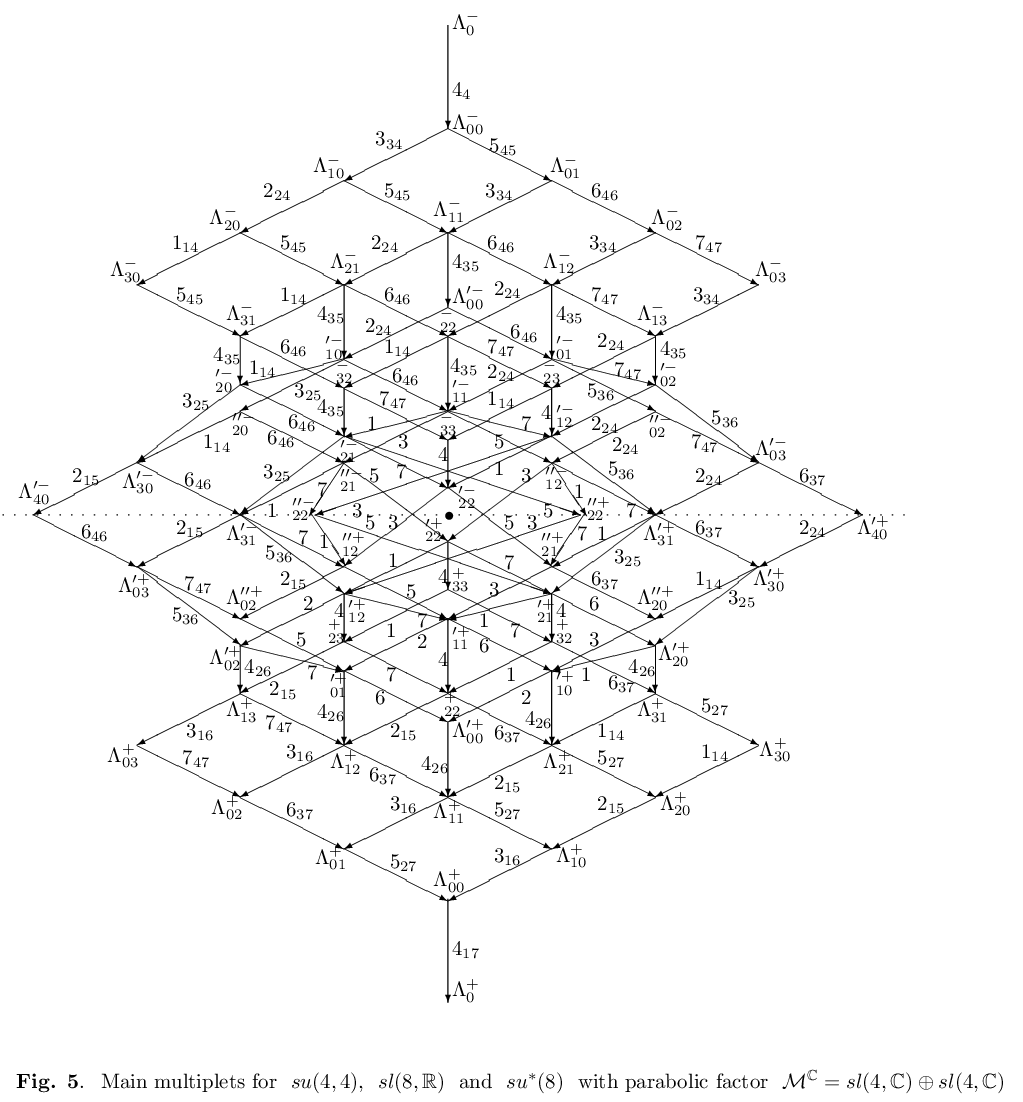}{20cm}


\section{The Lie algebras {\boldmath  $sp(p,r)$}}

Let ~$\cg = sp(p,r)$, ~$p\geq r$. It has maximal compact subalgebra
$\ck = sp(p) \oplus sp(r)$ and has discrete series representations
(as $\rank \ck = p+r = \rank \cg$). It has $r$ maximal parabolic
subalgebras with $\cm$-factors (cf. (9.8) from \cite{Dobinv}):
\eqn{cmsprrmax}
 \cm^{\rm max}_j ~=~ su^*(2j) \oplus sp(p-j,r-j)\  , ~~~1 \leq j \leq r\  \ee
with complexification:
\eqn{cmsprrmaz}
(\cm^{\rm max}_j)^\bac ~=~ sl(2j,\bbc) \oplus sp(p+r-2j,\bbc)\ . \ee

We would like to match this algebra with the appropriate conformal
Lie algebra, namely, with ~$sp(n,\bbr)$. It was considered in
\cite{Dobspn} with $\cm$-factor: ~$\cm' = sl(n,\bbr)$~  with
 complexification ~$\cm'^\bac = sl(n,\bbc)$.
Obviously, the latter can match \eqref{cmsprrmaz} only if ~$n$~ is even and
~$p=r=j=n/2$.  Thus, we shall consider
\eqnn{spprr} \cg ~&=&~ sp(r,r) \ , \\
\cm ~&=&~ su^*(2r) \ , \nn\\
\cm^\bac ~&=&~ sl(2r,\bbc) \ .\nn \eea

 The signature of the ERs of $\cg$   is:
\eqn{sgnspn}   \chi ~=~ \{\, n_1\,, \ldots,\, n_{2r-1}\, ;\, c\, \} \ ,
\qquad n_j \in \bbn\ ,  \qquad
 c = d- r - \ha \ . \ee

 The Knapp-Stein  restricted Weyl reflection  acts
as follows:
  \eqnn{kstspn} && G  ~:~ \cc_\chi ~ \llr ~ \cc_{\chi'} \ ,
  \\ &&\chi' ~=~
\{\, (n_1,\ldots,n_{2r-1})^* \,;\, -c\, \} \ , \qquad
(n_1,\ldots,n_{2r-1})^* ~\doteq~ (n_{2r-1},\ldots,n_{1})  \nn \eea

In terms of an orthonormal basis $\veps_i\,$, ~$i=1,\ldots,n$, the
 positive roots of $sp(2r,\bbc)$ are:  \eqn{pospn} \D^+ = \{ \veps_i \pm \veps_j, ~~1 \leq
i <j \leq 2r; \quad 2\veps_i, ~~1 \leq i \leq 2r\} \ , \ee  the simple
roots are: \eqn{simpspn}  \pi = \{\a_i = \veps_i - \veps_{i+1}, ~1 \leq
i \leq 2r - 1; \quad \a_{2r} = 2\veps_{2r}\}  \ , \ee
the positive non-compact roots are:
\eqn{posncspn} \b_{ij} ~\equiv~ \veps_i
+\veps_j,\ ,   ~~ 1 \leq i \leq j \leq 2r \ , \ee
the Harish-Chandra parameters:
 $m_\b \equiv (\L+\r, \b )$ for  the noncompact roots  are:
 \eqnn{hclab} m_{\b_{ij}} ~&=&~
\Big( \sum_{s=i}^{2r} + \sum_{s=j}^{2r} \Big)   m_s \ , \qquad i<j \ ,\\
m_{\b_{ii}} ~&=&~  \sum_{s=i}^{2r}    m_s  \nn\eea

 The correspondence between the signatures $\chi$ and
the highest weight $\L$ is:
\eqn{corspn}  n_i
= m_i \ , \quad  c ~=~ -\ha (m_\ta + m_{2r}) ~=~  -\,\ha(   m_1+\cdots +
m_{2r-1} + 2m_{2r}  ) \ee  where ~$\ta ~=~ \b_{11}$~ is the highest root.

The number of ERs in the corresponding multiplets is according to \eqref{multi}:
\eqn{multisp} \frac{\vr W(\cg^\bac,\ch^\bac)\vr}{\vr
W(\cm^\bac,\ch_m^\bac)\vr} ~=~ \frac{\vr W(sp(2r,\bbc))\vr}{\vr W(sl(2r,\bbc))\vr}
~=~ \frac{2^{2r}(2r)!}{ ((2r)!)} ~=~ 2^{2r} \ee
(which was given for $sp(n,\bbr)$ in \cite{Dobspn}).

Below we give pictorially the multiplets for ~$sp(r,r)$~ for ~$r=1,2$,
valid also for ~$sp(2r,\bbr$). (The case $r=3$, together with the reduced
multiplets and  $sp(5,\bbr$
are given in \cite{Dobspn}.)

In fact,  the case ~$r=1$~
is known long time as  ~$sp(1,1)\cong so(4,1)$, cf. \cite{DMPPT},
then later as $sp(2,\bbr)\cong so(3,2)$, cf. \cite{Dobso},   as we recalled already
in the previous section on $so(p,q)$ algebras.  We present it here using
 the new diagram look which we already used in the previous Section.
 Thus, in Figure 6 we give the case $r=1$, where the Knapp-Stein symmetry is
w.r.t. to the bullet in the middle of the figure.  Thus, it is seen that
the action of the differential operator indexed by ~$1_{12}$~ is the same
as the Knapp-Stein operator from ~$\L'^-$~ to ~$\L'^+$, so that the latter
operator degenerates as discussed in Section 1. Then in   Figure 7 we
give the diagram Figure 2 for the special case  $h=1$,
stressing that both Figures 6 and 7 have the same content.

Finally, in Figure 8 we give the case ~$r=2$.

\np

\fig{}{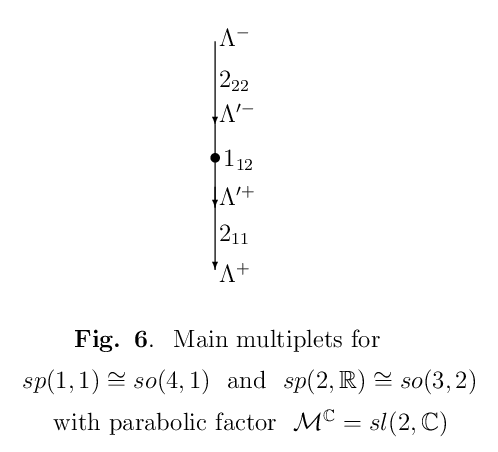}{8cm}

\fig{}{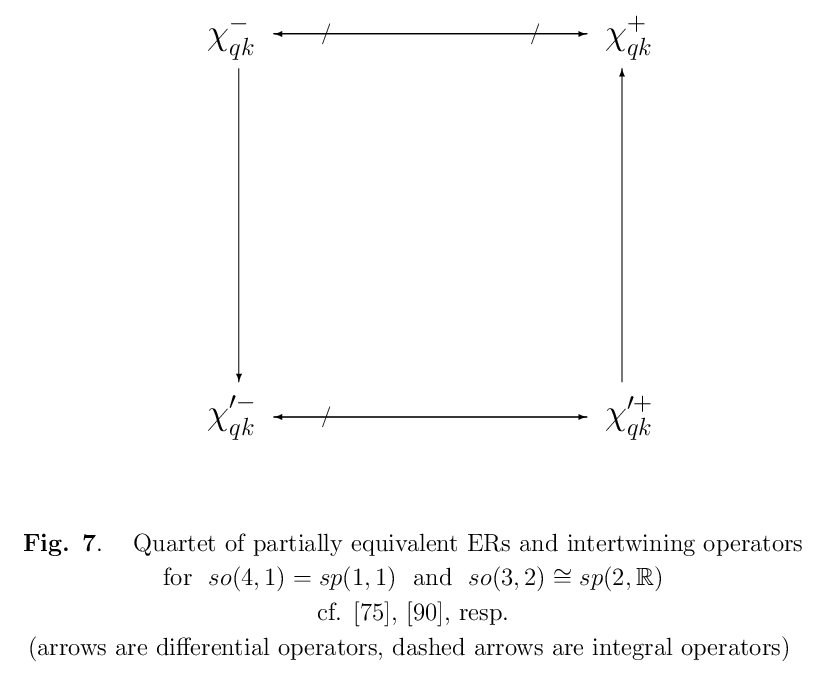}{11cm}

\fig{}{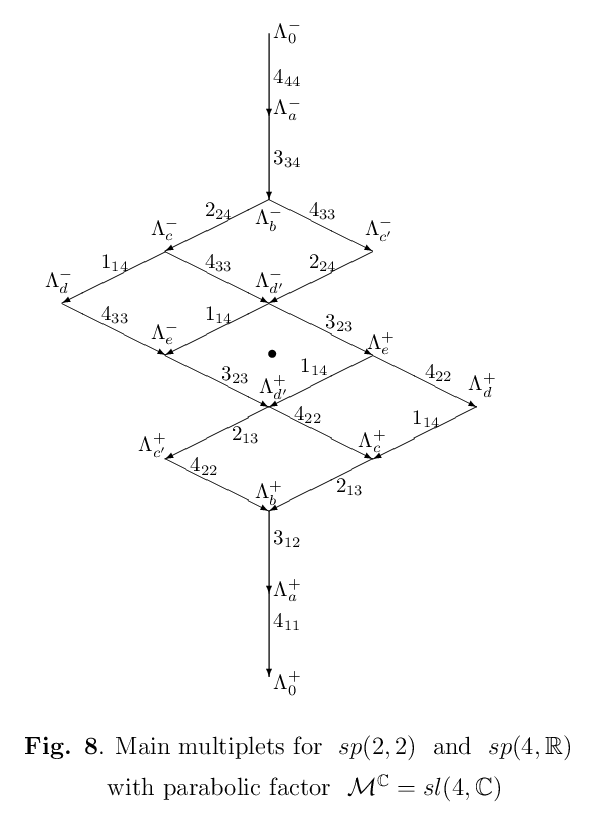}{13cm}

\np

\section{The non-compact Lie algebra {\boldmath $E_{7(7)}$}}

\nt Let ~$\cg=E_{7(7)}$. This is the split real form of ~$E_7$~
which is denoted also as ~$E'_7$~ or ~$EV$.   The maximal compact
subgroup is ~$\ck \cong su(8)$. This algebra has discrete series
representations (as $\rank \cg=\rank \ck$).

It has the following Dynkin-Satake diagram (same as for $E_7$)\cite{Sata}: \eqn{satseva}
\downcirc{{\a_1}} \riga\downcirc{\a_3} \riga
\downcirc{{\a_4}}\kern-8pt\raise11pt\hbox{$\vert$}
\kern-3.5pt\raise22pt\hbox{$\circ {{\scriptstyle{\a_2}}}$}
\riga\downcirc{{\a_5}} \riga\downcirc{{\a_6}}
\riga\downcirc{{\a_7}} \ee

The real algebra $E_{7(7)}$ has seven maximal parabolics which are obtained by
deleting one node as explained in \cite{Dobinv}. We
choose the one which is most suitable w.r.t. the maximal compact subgroup ~$\ck=su(8)$,
as will become clear below.
This parabolic is obtained by deleting
the root ~$\a_7$~ from the Dynkin-Satake diagram \eqref{satseva}, i.e., we shall use as $\cm$-factor
$E_{6(6)}\,$ (the split real form of $E_6$).

Thus, our  ~{\it maximal} parabolic is
\eqn{maxpar:77} \cp=\cm\oplus\ca\oplus\cn \ , \qquad
\ca\cong so(1,1)\ ,   ~~\cm\cong E_{6(6)}\ ,   ~~\dim_\bbr\,\cn = 27\ ,\ee
  cf. (11.17) of \cite{Dobinv}.

We label   the signature of the ERs of $\cg$   as follows:
\eqn{sgnde}  \chi ~=~ \{\, n_1\,, \ldots,\, n_{6}\,;\, c\, \} \ ,
\qquad n_j \in \bbn\ , \quad c = d- 9 \ee where the last entry of ~$\chi$~ labels
the characters of $\ca\,$, and the first $6$ entries are labels of
the finite-dimensional nonunitary irreps of $\cm\,$, (or of the
finite-dimensional unitary irreps of the compact ~$e_6$).

Further, we need the root system of the complex algebra ~$E_7\,$.
With Dynkin diagram enumerating the simple roots ~$\a_i$~ as in
\eqref{satseva}, the positive roots are:\nl first there are ~21~
roots forming the positive root system ~of ~$sl(7)$~ (with simple roots ~
$\a_1,\a_3,\a_4,\a_5$, $\a_6,\a_7\,$), ~~then ~21~ positive roots which are
positive roots of the ~$E_6$~ subalgebra including the non-$sl(7)$ root $\a_2$, and finally
the following ~21~ roots including the non-$E_6$ root ~$\a_7$~:
\eqnn{satsevy}
&&\a_2+\a_4+\a_5+\a_6+\a_7\ , ~~  \a_2+\a_3+\a_4+\a_5+\a_6+\a_7\ ,   \\
&&\a_1+ \a_2+\a_3+\a_4+\a_5+\a_6+\a_7 \ , \nn\\
&&   \a_2+\a_3+2\a_4+\a_5+\a_6+\a_7\ ,
~~\a_1+\a_2+\a_3+2\a_4+\a_5+\a_6+\a_7\ ,  \nn\\
&&   \a_2+\a_3+2\a_4+2\a_5+\a_6+\a_7\ ,
~~\a_1+\a_2+2\a_3+2\a_4+\a_5+\a_6+\a_7\ ,\nn\\
&& \a_1+\a_2+\a_3+2\a_4+2\a_5+\a_6+\a_7\ ,\nn\\
&&\a_1+\a_2+2\a_3+2\a_4+2\a_5+\a_6+\a_7\ ,\nn\\
&& \a_1+\a_2+2\a_3+3\a_4+2\a_5+\a_6+\a_7\ , \nn\\
&&
\a_1+2\a_2+2\a_3+3\a_4+2\a_5+\a_6+\a_7\ ,\nn\\
&&   \a_2+\a_3+2\a_4+2\a_5+2\a_6+\a_7\ ,\nn\\
&& \a_1+\a_2+\a_3+2\a_4+2\a_5+2\a_6+\a_7\ , \nn\\
&& \a_1+\a_2+2\a_3+2\a_4+2\a_5+2\a_6+\a_7\ ,\nn\\
&& \a_1+\a_2+2\a_3+3\a_4+2\a_5+2\a_6+\a_7\ , \nn\\
&& \a_1+2\a_2+2\a_3+3\a_4+2\a_5+2\a_6+\a_7\ ,\nn\\
&& \a_1+\a_2+2\a_3+3\a_4+3\a_5+2\a_6+\a_7\ , \nn\\
&& \a_1+2\a_2+2\a_3+3\a_4+3\a_5+2\a_6+\a_7\ ,\nn\\
&&  \a_1+  2\a_2+2\a_3+4\a_4+3\a_5+2\a_6+\a_7\ ,\nn\\
&& \a_1+2\a_2+3\a_3+4\a_4+3\a_5+2\a_6+\a_7\ ,\nn\\
&&  2\a_1+  2\a_2+3\a_3+4\a_4+3\a_5+2\a_6+\a_7 ~=~ \tilde{\a}\ ,
\nn \eea
where ~$\ta$~ is the highest root of the $E_7$ root system.

 The differential
intertwining operators that give the multiplets correspond to  the noncompact
roots, and since we shall use the latter extensively, we introduce more
compact notation for them. Namely, the non-simple roots  will be
denoted in a self-explanatory way as follows:
\eqnn{nota}
&&\a_{ij} ~=~ \a_i + \a_{i+1} + \cdots + \a_j \ ,
~~~\a_{i,j} ~=~ \a_i  + \a_j\ , ~~~i < j \ ,\\
&&\a_{ij,k} ~=~ \a_{k,ij} ~=~\a_i + \a_{i+1} +\cdots + \a_j +\a_k\ , ~~~i< j \ ,
\nn\\
&& \a_{ij,km} ~=~ \a_i + \a_{i+1} +\cdots + \a_j +\a_k+ \a_{k+1} +\cdots +\a_m\ ,\nn\\
&&  \qquad \qquad\qquad\qquad
~~~i< j \ , ~~k<m \ ,
\nn\\
&& \a_{ij,km,4} ~=~ \a_i + \a_{i+1} + \cdots + \a_j +\a_k+ \a_{k+1} +
\cdots +\a_m+\a_4\ , \nn\\
&&  \qquad \qquad\qquad\qquad~~~i< j \ , ~~k<m \ , \nn\eea
i.e., the non-compact roots will be written as:
\eqna{satsevyy}
 &&\a_7 \ , ~~\a_{67}\ , ~~\a_{57}\ , ~~\a_{47}\ , ~~\a_{37}\ , ~~\a_{1,37}\ , \\
 &&\a_{2,47}\ , ~~  \a_{27}  \ , ~~
\a_{17}  \ , ~~  \a_{27,4}  \ ,  ~~  \a_{17,4}  \ ,
~~  \a_{27,45}  \ , ~~\\ && \a_{17,34}  \ , ~~  \a_{17,45}  \ , ~~  \a_{27,46}
  \ ,~~  \a_{17,35}  \ , ~~  \a_{17,46}  \ ,
~~  \a_{17,36}  \ ,\nn\\ &&
 \a_{17,35,4}  \ , ~~  \a_{17,25,4}  \ ,
~~  \a_{17,36,4}  \ , ~~  \a_{17,26,4}  \ ,\nn\\ &&
 \a_{17,36,45}  \ ,
~~  \a_{17,26,45}  \ ,
~~  \a_{17,26,45,4}  \ ,
~~  \a_{17,26,35,4}  \ ,
~~  \a_{17,16,35,4}    ~=~ \tilde{\a}\ , \nn\eena
where the first six roots in (\ref{satsevyy}a)
are from the $sl(7)$ subalgebra, and the 21 in (\ref{satsevyy}b)
are those from  \eqref{satsevy}.

Further, we give the correspondence between the signatures $\chi$
and the highest weight $\L$.  The connection is through the Dynkin
labels \eqref{dynk}       ~$m_i$, $i=1,\ldots,7$, and is given
 explicitly by:
  \eqnn{rela} && n_i = m_i \ , \quad i=1,\ldots,6 \ , \\
&&  c = -\ha (m_\ta + m_7) =  -\ha(   2m_1+2m_2 +
3m_3 + 4m_4 + 3m_5 + 2m_6  + 2m_7)  \nn \eea

Here we note that the simple root system of the $su(8)$ compact subalgebra of ~$E_{7(7)}$,
or equivalently, of the $sl(8)$ subalgebra of $E_7\,$, is given by the
~$sl(7)$~ simple roots plus the highest root ~$\has$~ of the $E_6$ subalgebra:
\eqn{seight} \a_1,~\a_3,~\a_4,~\a_5,~ \a_6,~\a_7, ~
\has ~=~ \a_1+ 2\a_2+2\a_3+3\a_4+2\a_5+\a_6 \ee
Indeed, it is easy to check that:
$$(\a_i,\has) = 0 ,\quad i=1,3,4,5,6, \qquad  (\a_7,\has) = -1 \ .$$

Now we should connect our considerations with the case of another real form of $E_7\,$,
namely, the Lie algebra $E_{7(-25)}$, cf. \cite{Dobeseven}. In that paper we chose
as maximal parabolic ~$\cp' = \cm' \oplus \ca' \oplus \cn'$,
 where ~$\cm' \cong E_{6(-26)}$, ~$\dim_\bbr\,\cn = 27$,   cf. (11.24) of
\cite{Dobinv}.

Since the algebras $E_{7(7)}$ and $E_{7(-25)}$ are parabolically related
they have the   same signatures,
and thus the same main multiplets.

The number of ERs in the corresponding main multiplets is according to \eqref{multi}:
\eqn{multisev} \frac{\vr W(\cg^\bac,\ch^\bac)\vr}{\vr
W(\cm^\bac,\ch_m^\bac)\vr} ~=~ \frac{\vr W(E_7)\vr}{\vr W(E_6)\vr}
~=~ \frac{2^{10}\,3^4\,5.7}{2^7\,3^4\,5} ~=~ 56  \ee
(which was given for $E_{7(-25)}$ in \cite{Dobeseven}).

Below we give the main multiplets valid for both algebras in Figure 9.
For reduced multiplets cf. \cite{Dobeseven}.

\np\thispagestyle{empty}

\fig{}{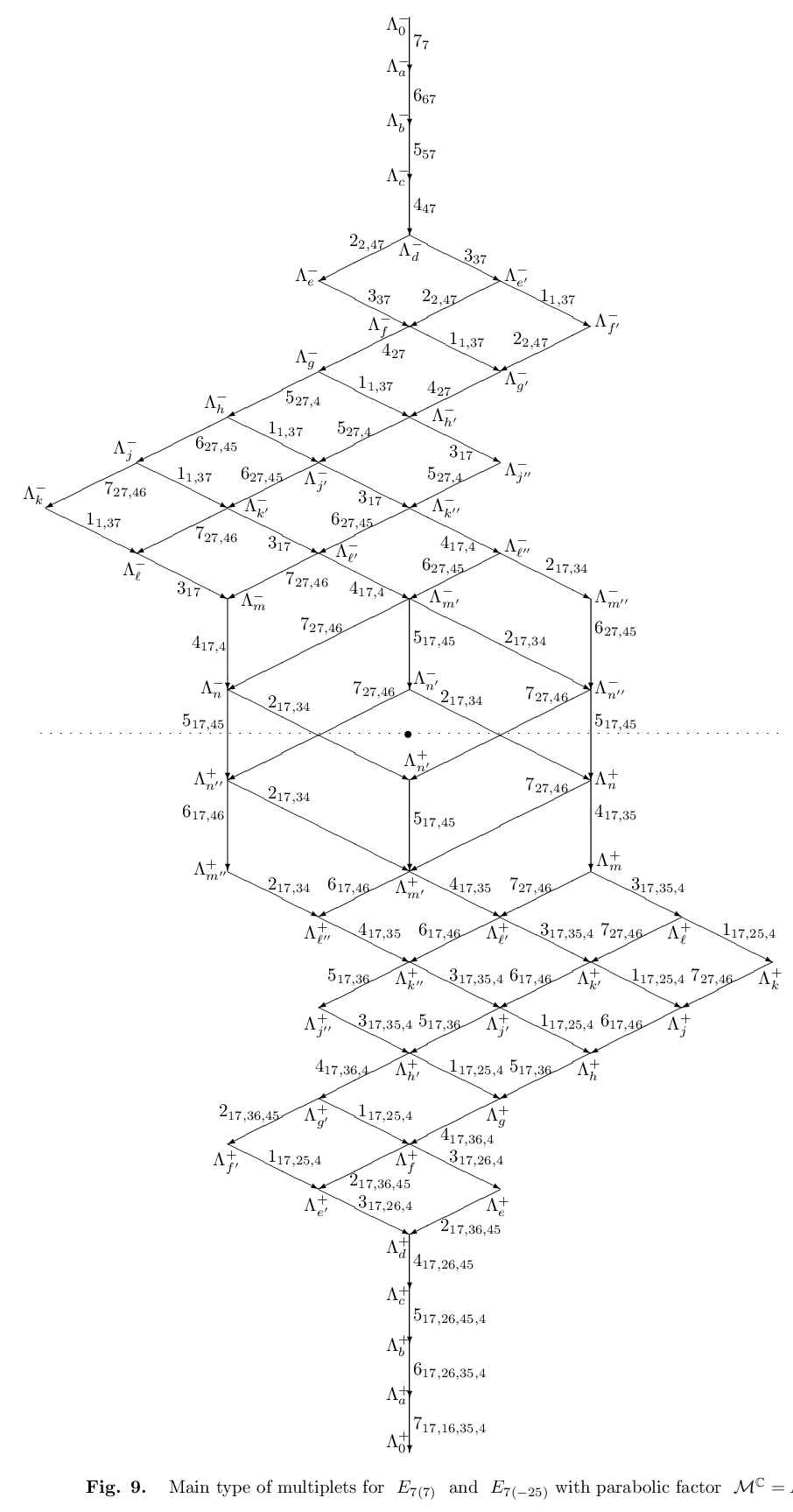}{13cm}

\np

\section{Two real forms of {\boldmath $E_6$}}

\subsection{The  Lie algebra {\boldmath  $E_{6(6)}$}}

Let ~$\cg ~=~ ~E_{6(6)}\,$. This is the split real form of ~$E_6$~
denoted  also as    ~$E'_6\,$ ~or ~$E_{\rm I}$.
The maximal compact subgroup is ~$\ck \cong sp(4)$.
This real form does not have discrete series representations
(as $\rank \cg \neq \rank \ck$).

 We use the following Dynkin-Satake diagram (same as for $E_6$):
\eqn{satsixa}\downcirc{{\a_1}} \riga\downcirc{\a_3} \riga
\downcirc{{\a_4}}\kern-8pt\raise11pt\hbox{$\vert$}
\kern-3.5pt\raise22pt\hbox{$\circ {{\scriptstyle{\a_2}}}$}
\riga\downcirc{{\a_5}} \riga\downcirc{{\a_6}} \ee

The real algebra $E_{6(6)}$ has four   maximal parabolics which are obtained by
deleting one node as explained in \cite{Dobinv}. (Note that deleting node 1 or node 6 produces
the same parabolic, same for deleting node 3 or node 5.) We choose the parabolic
obtained by deleting node 2.

Thus, the ~{\it maximal} parabolic is
\eqn{maxpareI}
\cp=\cm\oplus\ca\oplus\cn \ ,\qquad
\ca\cong so(1,1)\ ,   ~~\cm\cong sl(6,\bbr)\ ,   ~~\dim_\bbr\,\cn = 21\ ,\ee
 cf. (11.4) of \cite{Dobinv}.

\subsection{The  Lie algebra {\boldmath  $E_{6(2)}$}}

Let ~$\cg ~=~ E_{6(2)}\,$.  This is another real form of ~$E_6$~
sometimes denoted as    ~$E''_6\,$, or ~$E_{\rm II}\,$.
The maximal compact subalgebra is ~$\ck \cong su(6)\oplus su(2)$.
This real form has discrete series representations.

The Satake diagram is:
\eqn{satsixaz} \underbrace{ \downcirc{{\a_1}} \riga \underbrace{ \downcirc{\a_3} \riga
\downcirc{{\a_4}}\kern-8pt\raise11pt\hbox{$\vert$}
\kern-3.5pt\raise22pt\hbox{$\circ {{\scriptstyle{\a_2}}}$}
\riga\downcirc{{\a_5}}} \riga\downcirc{{\a_6}} }  \ee

The real algebra $E_{6(2)}$ has four   maximal parabolics which are obtained by
deleting one node as explained in \cite{Dobinv} (taking into account $E_6$ symmetry as
in the previous case). We choose the parabolic
obtained by deleting node 2.

Thus, the ~{\it maximal} parabolic is
\eqn{maxpareII}
\cp=\cm\oplus\ca\oplus\cn\ , \qquad
\ca\cong so(1,1)\ ,   ~~\cm\cong su(3,3)\ ,   ~~\dim_\bbr\,\cn = 21\ ,\ee
  cf. (11.7) of \cite{Dobinv}.

\subsection{Representations and multiplets}

We note that the $\cm$-factors of the two real forms of $E_6$ discussed in
the previous subsections have the same complexification:
$$ sl(6,\bbr)^\bac ~=~ su(3,3)^\bac ~=~ sl(6,\bbc) $$
i.e., they are parabolically related and we can discuss them together.

The signature of the ERs of $\cg$  is:
$$ \chi = \{\, n_1\,, n_3\,,n_4\,,n_5\,,n_6\,;\, c \} \
, \quad c = d- \hel\ , $$
expressed through the Dynkin labels as:  $$
n_i = m_i \ , \quad - c ~=~ \ha m_\ta ~=~ $$$$\ha ( m_1+2m_2 + 2m_3 +
3m_4 + 2m_5 + m_6) $$

Further, we need the root system of the complex algebra ~$E_6\,$.
With Dynkin diagram enumerating the simple roots ~$\a_i$~ as in
\eqref{satsixa}, the positive roots are:\nl first there are ~15~
roots forming the positive root system of ~$sl(6)$~ (with simple roots~~
$\a_1,\a_3,\a_4,\a_5,\a_6$), ~~then
the following ~21~ roots including the non-$sl(6)$ root ~$\a_2$~:
 \eqnn{satsixz}
&&\a_2\ , ~~\a_2+\a_4\ , ~~\a_2+\a_3+\a_4\ , ~~\a_2+\a_4+\a_5\
, \\ &&\a_2+\a_3+\a_4+\a_5\ , ~~\a_1+\a_2+\a_3+\a_4\ ,
~~\a_2+\a_4+\a_5+\a_6\ , \nn\\ &&  \a_1+\a_2+\a_3+\a_4+\a_5\ ,
~~\a_2+\a_3+\a_4+\a_5+\a_6\ , ~~\a_2+\a_3+2\a_4+\a_5 ,\nn\\ &&
\a_1+\a_2+\a_3+\a_4+\a_5+\a_6 \ , ~~\a_1+\a_2+\a_3+2\a_4+\a_5\ ,\nn\\ &&
\a_2+\a_3+2\a_4+\a_5+\a_6\ ,  ~~\a_1+\a_2+\a_3+2\a_4+\a_5+\a_6\ ,\nn\\ &&
\a_1+\a_2+2\a_3+2\a_4+\a_5\ , ~~\a_2+\a_3+2\a_4+2\a_5+\a_6\ ,
\nn\\ &&  \a_1+\a_2+2\a_3+2\a_4+\a_5+\a_6\ ,
~~\a_1+\a_2+\a_3+2\a_4+2\a_5+\a_6\ , \nn\\ &&
\a_1+\a_2+2\a_3+2\a_4+2\a_5+\a_6\ ,\nn\\ &&
\a_1+\a_2+2\a_3+3\a_4+2\a_5+\a_6\ , \nn\\ &&
\a_1+2\a_2+2\a_3+3\a_4+2\a_5+\a_6 ~\equiv~ \ta \ ,\nn \eea where ~$\ta$~
is the highest root of the $E_6$ root system.

Relative to our parabolic subalgebra, the roots in \eqref{satsixz} are
non-compact, while the rest are compact. As before we introduce
 more condensed notation for the  noncompact roots:
  \eqnn{satsixzz}
&&\a_2\,,~ \a_{14}\,,~ \a_{15}\,,~\a_{16} \,,~  \a_{24}\,,~ \a_{25}\,,~\a_{26}  \cr &&
 \a_{2,4}\,,~ \a_{2,45}\,,~
 \a_{2,46}\,,~
  \a_{25,4}\,,~  \a_{15,4} \,,~\a_{26,4}  \cr
&&\a_{16,4} \,,~  \a_{15,34} \,,~ \a_{26,45} \,,~
\a_{16,34} \,,~ \a_{16,45}   \cr &&\a_{16,35}\,,~  \a_{16,35,4}\,,~
\a_{16,25,4} ~=~ \ta \  \nn
 \eea

Now we should connect our considerations with the case of another real form of $E_6\,$,
namely, the Lie algebra $E_{6(-14)}$, cf. \cite{Dobesix}. In that paper we chose
as maximal parabolic ~$\cp' = \cm' \oplus \ca' \oplus \cn'$,
 where ~$\cm' \cong su(5,1)$, ~$\dim_\bbr\,\cn = 21$,   cf. (11.21) of
\cite{Dobinv}.

Since both the algebras and the maximal parabolics have the same complexification,
this means that they are parabolically related,
thus, we  have the same non-compact roots, the same signatures,
and the same multiplets.  We show only the main multiplet in Figure 10, referring
to \cite{Dobesix} for the diagrams of reduced multiplets. The main multiplet has 70 members
and the figure has the standard $E_6$ symmetry, namely,
conjugation exchanging indices $1\llra 6$, $3\llra 5$.
The Knapp-Stein operators act pictorially as reflection w.r.t. the dotted line separating the
~$\ch^-{...}$~ members from the $\ch^+{...}$ members.
 Note that there are five cases when the embeddings correspond to the
highest root $\ta$~: ~~$V^{\L^-} \lra V^{\L^+}$, ~$\L^+ ~=~\L^-
-m_\ta\,\ta\,$. In these five cases the weights are denoted as:
~$\L^\pm_{k''}\,$, ~$\L^\pm_{k'}\,$, ~$\L^\pm_{\tk}\,$,
~$\L^\pm_{k}\,$, ~$\L^\pm_{k^o}\,$, then:   ~$m_\ta ~=~
m_1,m_3,m_4,m_5,m_6\,$, resp. We recall that Knapp-Stein operators
~$G^+$~ intertwine the corresponding ERs ~$\ct_\chi^-$ and
~$\ct_\chi^+$. In the above five cases the Knapp-Stein
operators ~$G^+$~ degenerate to differential operators as we discussed earlier.

\np\thispagestyle{empty}   \voffset -45mm

\fig{}{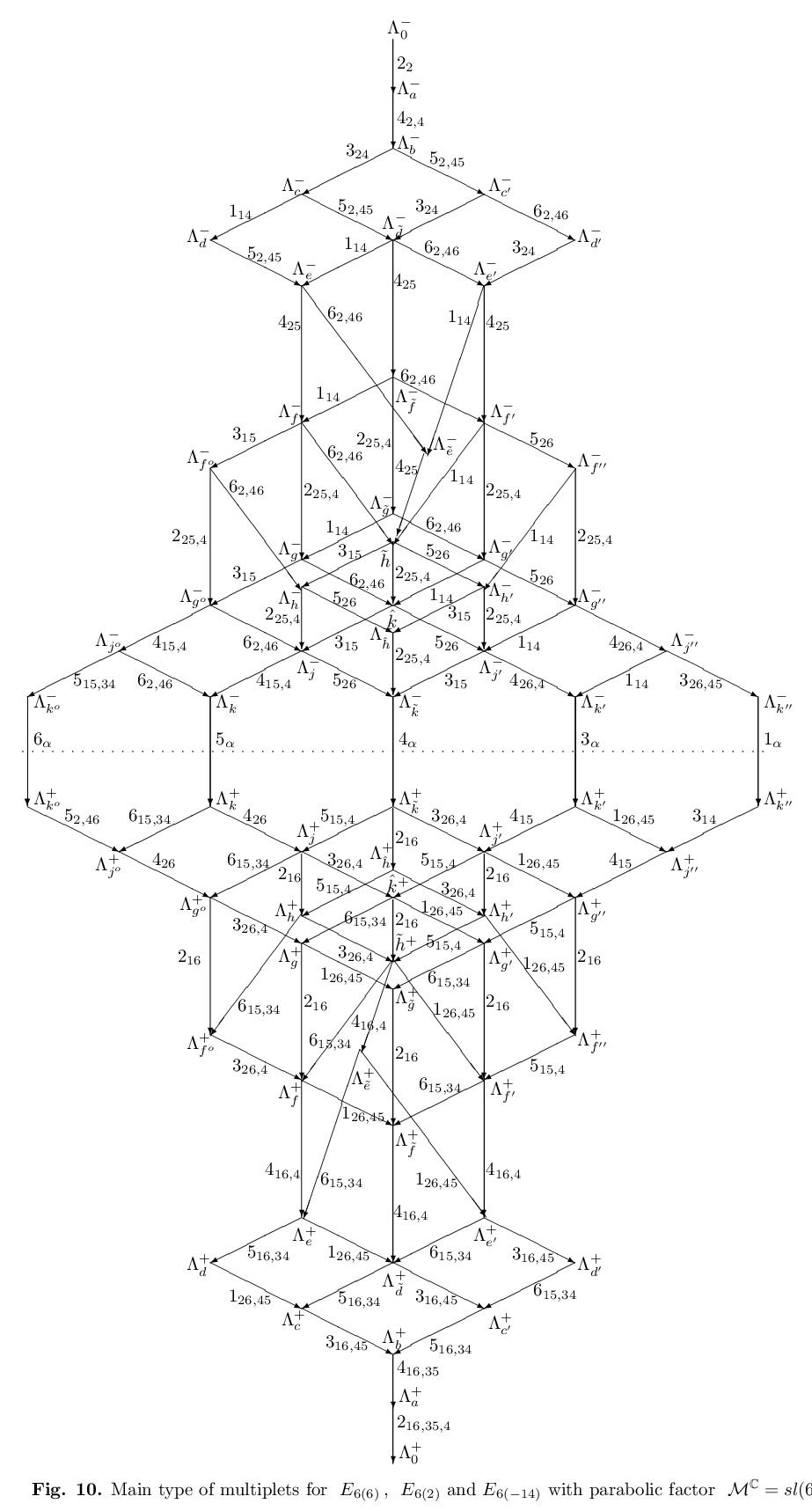}{155mm}

\np \voffset 0cm

\section{Summary and Outlook}

In the present paper we continued the project of systematic
construction of invariant differential operators for non-compact semisimple Lie groups.
Our aim in this paper was to extend our considerations beyond the
  class of algebras, which we call 'conformal Lie
algebras' (CLA).       For this we introduce the new notion of ~{\it
parabolic relation}~ between two non-compact semisimple Lie algebras
$\cg$ and $\cg'$ that have the same complexification and possess
maximal parabolic subalgebras with the same complexification. Thus,
we considered the algebras ~$so(p,q)$~ all of which are
parabolically related to the conformal  algebra ~$so(n,2)$~ with
~$p+q=n+2$, then the algebras ~$su^*(4k)$~ and ~$sl(4k,\bbr)$~
parabolically related to the CLA ~$su(2k,2k)$, then ~$sp(r,r)$~ as
parabolically related to the CLA ~$sp(2r)$ (of rank $2r$), then the
exceptional Lie algebra ~$E_{7(7)}$~ which is parabolically related
to the CLA ~$E_{7(-25)}\,$, finally the exceptional Lie algebras
~$E_{6(6)}$~ and ~$E_{6(2)}$~
 parabolically related to the hermitian symmetric case ~$E_{6(-14)}\,$.

We have given a formula for the number of representations in the main multiplets valid
for CLAs and all algebras that are parabolically related to them.
In all considered cases we have given the main multiplets of indecomposable elementary representations
including the necessary data for all relevant invariant differential operators.
In the case of $so(p,q)$ we have given  also the  reduced multiplets.
We note that the multiplets are given in the most economic way in pairs of
~{\it shadow fields}~ related by the Knapp-Stein restricted Weyl symmetry (and
the corresponding integral operators).

Finally, we should stress that the classification
of all invariant differential operators includes as special cases all possible ~{\it
conservation laws}~ and ~{\it conserved currents}, unitary or not.

We plan also to extend these considerations to the supersymmetric
cases and also to the quantum group setting. Such considerations are
expected to be very useful for applications to string theory and
integrable models. It is interesting to note that almost all of the
algebras that appear in Table 1 of \cite{FerrKaMa} are treated in
the present paper, though our motivations and approach are different
(see also \cite{FMOSTYZ}).

\vskip 5mm

\noindent {\bf Acknowledgments.} The author thanks S. Ferrara for
stimulating discussions. The author
thanks the Theory Division of CERN for hospitality during the course
of this work.
 This work was supported in part by the Bulgarian National Science Fund, grant DO 02-257.

\np

\end{document}